\documentclass[pra,twocolumn,showpacs,amsmath,amssymb]{revtex4}
\usepackage{amsfonts}
\usepackage{graphicx}
\usepackage{amsmath}
\usepackage{bm}
\usepackage{amssymb}
\usepackage{soul}
\usepackage{color}
 \usepackage[version=3]{mhchem}
\usepackage[colorlinks,bookmarks=false,citecolor=blue,linkcolor=red,urlcolor=blue]{hyperref}

\usepackage{multirow}

\begin{document}

\title{Quantum and Thermal Phase Transitions in a Bosonic Atom-Molecule Mixture in a Two-dimensional Optical Lattice}

\author{L. de Forges de Parny$^{1, 2}$ and V.G. Rousseau$^3$}  
\affiliation{$^1$ Laboratoire de Physique, CNRS UMR 5672,  \'Ecole Normale Sup\'erieure de Lyon, Universit\'e de Lyon, 46 All\'ee d'Italie, Lyon, F-69364, France}
\affiliation{$^2$ Physikalisches Institut, Albert-Ludwigs Universit\"{a}t Freiburg, Hermann-Herder Stra{\ss}e 3, D-79104, Freiburg, Germany}
\affiliation{$^3$ Physics Department, Loyola University New Orleans, 6363 Saint Charles Ave., New Orleans, Louisiana 70118, USA}

\date{\today}

\begin{abstract}
We study the ground state and the thermal phase diagram of a two-species Bose-Hubbard model, with U(1)$\times \mathbb{Z}_2$ symmetry,  describing atoms and molecules on a two-dimensional optical lattice interacting via a Feshbach resonance. Using quantum Monte Carlo simulations and mean field theory,  we show that the conversion between the two species, coherently coupling the atomic and molecular states, has a crucial impact on the Mott-superfluid transition and stabilizes an insulating phase with a gap controlled by the conversion term  -- \textit{the Feshbach insulator} -- instead of a standard Mott insulating phase. 
Depending on the detuning between atoms and molecules, this model exhibits three phases:  the Feshbach insulator, a molecular condensate coexisting with noncondensed atoms and a mixed atomic-molecular condensate.
Employing  finite-size scaling analysis, we observe  three-dimensional (3D) $XY$ (3D Ising) transition when U(1)  ($ \mathbb{Z}_2$) symmetry is broken whereas the transition is first-order when both U(1)  and $ \mathbb{Z}_2$ symmetries are spontaneously broken. The finite temperature phase diagram is also discussed. 
The thermal disappearance of the molecular superfluid leads to a Berezinskii-Kosterlitz-Thouless transition with unusual universal jump in the superfluid density. 
The loss of the  quasi-long-range coherence of the mixed atomic and molecular superfluid is more subtle since only atoms exhibit conventional Berezinskii-Kosterlitz-Thouless criticality.
We also observe a signal compatible with a  classical first-order transition between the mixed superfluid and the normal Bose liquid at low temperature.
\end{abstract}


\maketitle

\section{Introduction}

Ultracold atoms in optical lattices have opened new perspectives in several modern fields of physics, such as many-body  and condensed matter physics.
They offer  possibilities to study complex many-body systems \cite{Bloch_2008} and quantum phase transitions \cite{Sachdev_1999, Greiner_2002}
with a high degree of control.  
More interestingly, they are quantum simulators, giving access to the experimental implementation of models which are not easily realizable in any other physical contexts.
In all these applications, Feshbach resonances offer an invaluable tuning knob for controlling  the interaction between the atoms \cite{Chinetal2010}
and also give the possibility of coherently coupling different internal states of atoms.
As an example, an unbound state of two interacting atoms and a bound state -- hereafter called the \textit{molecular state} -- can be  brought into resonance by the application of
a magnetic field, thanks to the different magnetic moments  of the two states. 
 Therefore, ultracold atoms in optical lattices are also very suitable to study mixtures of
Bose-Einstein condensates (BECs), involving a coherent coupling \emph{\`a la} Josephson between pairs of atoms and molecules -- realizing  quantum-coherent chemical reactions.
The control of the effective scattering length of unbound atoms has led to the exploration of complex quantum many-body phases \cite{Greineretal2003, KetterleZ2008, RanderiaT2014, Winkleretal2006, Chinetal2010, Inouyeetal1998, Donley2002, Stengeretal1999}  and to tune transitions between them \cite{Jordensetal08, Deissleretal2010}, whereas the control of the 
coherent coupling between atoms and molecules has been exploited {e.g.}  to observe  atom-molecule Rabi oscillations \cite{Syassenetal2007, Buschetal1998, Olsenetal2009}.
Theoretically, the coherent coupling is at the basis of the prediction of a quantum phase transition between mixed atom-molecule and purely molecular condensates \cite{Radzihovskyetal2004,Romansetal2004,SenguptaD2005,deForges_Roscilde_2015,Radzihovskyetal2008,Ejima_2011,Bhaseenetal2012, Capponi_2007, Roux_2009}.
The case of bosonic atoms and molecules is all the more interesting since long-range phase coherence can be established in two dimensions  at zero temperature via Bose-Einstein condensation.
The asymmetric coherent coupling between atoms and molecules clearly leads to a complex interplay between atomic and molecular condensations and provides quantum phase transitions
which are not realized in the context of single species condensates. 
Even more striking is that the coherent coupling can destroy the phase coherence, leading to an insulating phase with a gap controlled by the conversion amplitude \cite{deForges_Roscilde_2015}.

In this paper we focus our attention on the case of a two dimensional (2D) atom-molecule mixture using quantum Monte Carlo  (QMC) simulations
and Gutzwiller mean field theory (MFT). Our goal is twofold: we elucidate the effect of the conversion term, leading to a rich and original ground state phase diagram 
and we unveil the thermal phase transitions, exhibiting an unusual Berezinskii-Kosterlitz-Thouless (BKT) transition.
Basically, we expect  mixed Mott insulator and superfluid phases, i.e., composed by both atoms and molecules, to appear due to the conversions. 
However, the effect of conversions on the phase transitions is more subtle and we report here a clear evidence that the phase coherence 
is destroyed when the conversion amplitude is increased.
Indeed, close to the resonance, the conversion term has a crucial impact on the Mott-superfluid transition for two particles per site: 
it enhances the insulator phase and also changes the nature of the transition, leading to a quantum first-order transition with a U(1)$\times \mathbb{Z}_2$
symmetry breaking. 
Even more interestingly, the phase located at the tip of the insulating lobe with two particles per site in the phase diagram does not correspond anymore with 
the definition of a Mott insulator phase, i.e an insulating phase with a particle-hole gap opened 
by (diagonal) repulsive interactions between the particles \cite{Mott_definition}.
Instead of this, the system adopts a Feshbach insulating phase where the  energy gap is controlled by the 
(off-diagonal) conversion term between atoms and molecules, keeping the interactions fixed.
Although the existence of this phase was previously reported in Ref.~\cite{deForges_Roscilde_2015}, 
our present study provides a reliable analysis of the quantum phase transitions,  completing the characterization of the phase diagram.
Finally, we study the thermal phase transitions. In two dimensions, Bose-Einstein condensation at finite temperature is impossible in the proper sense, 
leaving space to quasicondensation via a BKT transition \cite{Josebook}.  
The atomic and molecular conversions induce an asymmetric coupling between the phase of the atomic and molecular wave functions which couples the 
topological excitations (vortex-antivortex pairs) of both  fields. 
For positive detuning, this leads to an unusual BKT transitions when the quasi-long-range coherence of the mixed atomic and molecular superfluid is destroyed: 
only atoms exhibit conventional BKT criticality whereas molecules quasicondense in the same way as atom pairs condense.
Our QMC simulations are in agreement with  a previous study of an effective $XY$ coupled model, mimicking the atomic and molecular superfluid for positive detuning \cite{deForges2016}.
We complete here the picture by studying the thermal transition for negative detuning: 
the molecular condensate, coexisting with noncondensed atoms, exhibits conventional  BKT criticality but is found to be consistent with an
anomalous stiffness jump at the transition.\\

The paper is organized as follows:
The atom-molecule Hamiltonian  is presented in Sec.~\ref{section2}.
In Sec.~\ref{section3}, we discuss the mean field and quantum Monte Carlo approaches used to study the Hamiltonian. 
We also define the main observables of interest.
Section ~\ref{section4} is devoted to the discussion of the ground state phase diagram.
Quantum Monte Carlo calculations verify the qualitative
conclusions of the mean field theory, but provide
quantitatively accurate values for the phase boundaries and 
elucidate the universality classes of the quantum phase transitions.
Finally, in Sec.~\ref{section5}, we discuss the thermal phase diagram and 
the nonconventional  BKT transitions associated with the loss of molecular and atomic-molecular quasi-long-range coherence.
Conclusions and outlook are provided in Sec.~\ref{section6}.

\section{Atom-Molecule Hamiltonian}
\label{section2}

We consider spinless bosons with mass $m$ on a square optical lattice close to a narrow Feshbach resonance.
The system is described by a single-band Bose-Hubbard model with atomic and molecular bosons, 
coherently coupled via atom-atom interactions \cite{Dickerscheid2005}. 
The particles can hop between nearest neighboring sites, and their interactions are described by intraspecies and
interspecies onsite potentials. An additional term takes into account the conversion between two atoms and a molecule, and vice
versa.  The Hamiltonian of the system reads  \cite{Dickerscheid2005, Koehleretal2006, SenguptaD2005, Radzihovskyetal2004, Romansetal2004}
$\mathcal {  \hat H} =    \mathcal {   \hat T}+    \mathcal {  \hat P}+  \mathcal {  \hat C}$, where
\begin{eqnarray}
\label{Hamiltonian_atome_molecule_1}
 \mathcal {   \hat T} & = &  - \sum_{\langle i, j \rangle}  \left( t_a ~ a^\dagger_{i} a^{\phantom\dagger}_{j} + 
t_m ~m^\dagger_{i} m^{\phantom\dagger}_{j} + {\rm H.c.}  \right )~,\\
 \mathcal {   \hat P} &=& \sum_i \Big[ ~\frac{U_{a}}{2}  n^a_{ i} \left ( n^a_{i}-1\right )  
+  \frac{U_{m}}{2} n^m_{i} \left ( n^m_{i} -1\right )  \\
 \label{Hamiltonian_atome_molecule_2}
&+ &  \nonumber U_{am} n^a_{ i} n^m_{i}
 +  (U_{a}+\delta)    n^m_{i}   - \mu   \left( n^a_{i} +  2 n^m_{i} \right)\Big]~,  \\
\mathcal {  \hat C} &=&  g   \sum_{i}   \left (  m^\dagger_{i} a^{\phantom\dagger}_{i} a^{\phantom\dagger}_{i} 
+   a^{\dagger}_{i} a^{\dagger}_{i} m^{\phantom\dagger}_{i}   \right )~.
\label{Hamiltonian_atome_molecule}
\end{eqnarray}
The $  \mathcal {   \hat T}$ operator corresponds to the kinetic energy for hopping between nearest neighboring 
sites $\langle i, j \rangle$ defined on a $L\times L$ square lattice with periodic boundary conditions. 
Here $t_a$ and $t_m$ are respectively the tunneling amplitudes for the atoms and the molecules.
The $a^ \dagger_{i}$ and $a^{\phantom\dagger}_{i}$ ($m^ \dagger_{i}$ and $m^{\phantom\dagger}_{i}$) operators 
are bosonic creation and annihilation operators of atoms (molecules) on site 
$i$. $n^a_{i}= a^{\dagger}_{i}  a^{\phantom\dagger}_{i}$ and $n^m_{i}= m^{\dagger}_{i}  m^{\phantom\dagger}_{i}$ are the corresponding number operators. 
The $ \mathcal {  \hat P}$ operator contains the intraspecies  (interspecies) interactions with repulsive cost $U_a$ and $U_m$ ($U_{am}$), as well as the chemical potential term; 
in particular it contains the detuning term $\delta$ {\color{black} (controlled experimentally by a magnetic field \cite{Chinetal2010})}, 
which brings the state of two atoms and a molecule in and out of resonance on each site,  $\delta<0$ ($\delta >0$) corresponding to the molecular (atomic) side of the resonance. 
Finally the $ \mathcal {  \hat C} $ operator is the conversion term, which  coherently transforms a pair of atoms into a molecule and vice
versa  \cite{Radzihovskyetal2008}.
The conversion rate between atoms and molecules, $g$, is obtained via the solution of the scattering problem for two atoms in a parabolic potential \cite{Buschetal1998}. 
Following Ref.\cite{Syassenetal2007}, the parameter $g$, calculated by assuming a single harmonic potential, which is a good approximation for a deep optical lattice,  is given by
\begin{eqnarray}
g=  \left [ \frac{4  \pi  \hbar^2   a_{\rm{bg}}    \Delta \mu \Delta B    }{m (\sqrt{2 \pi} a_{\rm{ho}})^3} 
 \left ( 1+0.490 \frac {a_{\rm{bg}}}{a_{\rm{ho}}}  \right ) \right ]^{\frac{1}{2}}~,
\label{terme_g}
\end{eqnarray}
where $a_{\rm{ho}} =\sqrt{\hbar/ m \omega}$ the harmonic oscillator length, 
$a_{\rm{bg}}$ the background scattering length of the atoms,
$\Delta B$ the width of the Feshbach resonance,
and  $\Delta \mu$ the difference between the magnetic moments of an entrance-channel atom pair and a 
closed-channel molecule.  
The model described by the Hamiltonian $\mathcal {\hat H}$  remains realistic as long as $t_{a}, t_{m}, |V_{aa}|   << \hbar \omega$, 
with $V_{aa}$  the nonresonant atom-atom interaction and $\hbar \omega$ the energy splitting of the on-site optical lattice potential; 
see Ref.~\cite{Dickerscheid2005} for the conditions of applicability of this model and for the derivation of the Hamiltonian 
Eqs.~(\ref{Hamiltonian_atome_molecule_1} -- \ref{Hamiltonian_atome_molecule}) from a microscopic model.
Furthermore, the  validity of the single-band approximation requires $\sqrt{2} a_{\rm{bg}}   \Delta \mu \Delta B  / a_{\rm{ho}} \ll \hbar \omega$.
In other words,  the single-band approximation is is well respected for a narrow Feshbach resonance, 
e.g. $\Delta B =15 $ mG for $^{87}Rb$ near 414 G \cite{Syassenetal2007}.

The Hamiltonian $\mathcal {  \hat H}$ has the symmetry U(1)$\times \mathbb{Z}_2$, associated with the mass conservation in the mixture (U(1) symmetry), 
times the Ising $\mathbb{Z}_2$ symmetry in the phase relationship between atoms and molecules. 
As we discuss later, this emergent Ising symmetry, arising from the asymmetric nature of the atom-molecule coupling, is crucial for the understanding of the phase diagram.
A theoretically sound treatment requires one to take into account the full many-body physics of the Hamiltonian $\mathcal {  \hat H}$ which
 is  a rather hard task, given the large number of  parameters ($t_a, t_m, U_a, U_m, U_{am}, \mu, \delta, g$).
In order to simplify our study, we treat the parameters of the model as free parameters and
we consider symmetric parameters for atoms and molecules, leading to 
\begin{eqnarray}
\label{symmetric_params_1}
\begin{aligned}
t &\equiv& t_a=t_m~,\ \ \ \  \ \ \ \   \ \ \    \\
U &\equiv& U_a=U_m=U_{am}~,
\label{symmetric_params}
\end{aligned}
\end{eqnarray}
reducing the number of parameters to four independent parameters only : $t/U, \delta/U, \mu/U$ and $g/U$, where  
we choose the hopping parameter $t=1$ in order to set the energy scale.
A realistic scenario requires the calculation of the parameters from the microscopic Hamiltonian using the Wannier functions.
Nevertheless, since the qualitative aspects of the phase diagram do not depend on the precise values of $g$, $U_a/U_m$ and $t_a/t_m$, 
our choice of Eq.~(\ref{symmetric_params_1}) is indeed relevant \cite{SenguptaD2005} and 
captures the physics arising from the conversion term Eq.~(\ref{Hamiltonian_atome_molecule}), which is demonstrated in Ref.\cite{deForges_Roscilde_2015}.

The above atom-molecule Hamiltonian has been mainly studied using mean-field theory \cite{Radzihovskyetal2004, Romansetal2004, Radzihovskyetal2008}. 
The quantum phase transitions exhibited by the model have been extensively studied in one dimension  \cite{RousseauD2008, EckholtR2010, Ejima_2011, Bhaseenetal2012} whereas  
few studies have examined this question in two dimensions   \cite{SenguptaD2005, deForges_Roscilde_2015}. 
Here we numerically investigate this Hamiltonian in two dimensions, by using  exact QMC simulations based on the stochastic Green function algorithm \cite{SGF, directedSGF} and Gutzwiller mean field 
approach. We investigate both the quantum and thermal phase transitions.

\section{Methods}
\label{section3}

To capture the zero-temperature physics of the model, we use both the QMC method and the MFT approach.
While the QMC simulations become rather demanding for the calculation of the phase diagram, MFT allows for a rapid reconstruction of the phase boundaries, 
which turns out to be essential given the rich structure of the phase diagram - containing different critical and multicritical points. 

\subsection{Gutzwiller mean-field approach}
\label{section3_subA}

Although the mean field approximation does not give quantitatively accurate values for the phase boundaries, 
the mean-field phase diagram of a bosonic coupled mixture is in  good agreement with QMC simulations in two dimensions  at zero temperature \cite{deforges11, deforges13}
but fails at finite temperature \cite{deforges12}.
We use a mean-field formulation based on a decoupling approximation which decouples the
hopping  term  to  obtain  an  effective  one-site  problem. 
Introducing the atomic (molecular) superfluid order parameter  $\psi_a \equiv \langle
a^\dagger_{i}\rangle = \langle a_{i}\rangle$ ($\psi_m \equiv \langle
m^\dagger_{i}\rangle = \langle m_{i}\rangle$), we replace the creation and destruction operators on
site $i$ by their mean values $\psi_a$ and $\psi_m$. 
Since  we  are interested in equilibrium states, the order parameters can be chosen to be real.  
 Using this ansatz, the kinetic
energy terms, which are nondiagonal in boson creation
and destruction operators, are decoupled as
\begin{eqnarray}
\nonumber
a^{\dagger}_{i}   a^{\phantom\dagger}_{j} &=&
  \left (a^{\dagger}_{i}-\psi_a  \right )
  \left (a^{\phantom\dagger}_{j}- \psi_a  \right ) +   (a^{\dagger}_{i}  + a^{\phantom\dagger}_{j})  \psi_a - \psi_a^2\\
&& \simeq (a^{\dagger}_{i}  + a^{\phantom\dagger}_{j})  \psi_a - \psi_a^2   ~.
\label{mfapprox}
\end{eqnarray}
The same approximation applies for the molecules.
The Hamiltonian $\mathcal {  \hat H}$ is rewritten as a sum over local terms 
$\mathcal {  \hat H} =  \sum_i \mathcal {  \hat H}^{MF}_i$ where
\begin{eqnarray}
\nonumber
\mathcal {  \hat H}^{MF}_i &=& -4 t_a  (a^{\dagger}_{i}  + a^{\phantom\dagger}_{i})  \psi_a -4 t_m (m^{\dagger}_{i}  + m^{\phantom\dagger}_{i})  \psi_m\\
\nonumber
&& + 4 t_a \psi_a^2 + 4 t_m \psi_m^2 + U_{am} n^a_{ i} n^m_{i}\\
\nonumber
&& +  \frac{U_{a}}{2}  n^a_{ i} \left ( n^a_{i}-1\right )  
+  \frac{U_{m}}{2} n^m_{i} \left ( n^m_{i} -1\right )  \\
\nonumber
&& +  (U_{a}+\delta)    n^m_{i}   - \mu   \left( n^a_{i} +  2 n^m_{i} \right)\\
&&+ g    \left (  m^\dagger_{i} a^{\phantom\dagger}_{i} a^{\phantom\dagger}_{i} 
+   a^{\dagger}_{i} a^{\dagger}_{i} m^{\phantom\dagger}_{i}   \right )~.
\label{Hamiltonian_MF}
\end{eqnarray}
The mean field Hamiltonian Eq.~(\ref{Hamiltonian_MF})  can be easily
diagonalized numerically in a finite occupation-number  basis $\{ |n_a, n_m \rangle \}$, 
with the truncation $n_a^{max}=n_m^{max}=10$, 
and then minimizing
the lowest eigenvalue with respect to the order parameters
$\psi_a$ and $\psi_m$. 
This gives the order parameters of the ground state
and its eigenvector $|\Psi_{0}^{MF}  \rangle$.
At zero temperature, the system is in a Bose-Einstein condensate  phase if at  least  one  of  the  order  parameters is  nonzero
and is, otherwise, in an insulating phase.
The atomic and molecular condensate fraction is given by 
\begin{eqnarray}
C^{MF}_{\alpha}=|\psi_\alpha|^2~, 
\label{MF_condensate_fraction}
\end{eqnarray}
with $\alpha=a, m$.
The atomic and molecular densities are respectively  defined by 
\begin{eqnarray}
\nonumber
\rho_{a} &=&\langle \Psi_{0}^{MF}  | a^{\dagger}  a |\Psi_{0}^{MF}  \rangle~,\\
\rho_{m}&=&\langle \Psi_{0}^{MF} | m^{\dagger}  m |\Psi_{0}^{MF}  \rangle~.
\label{MF_density}
\end{eqnarray}
Finally, the compressibility is given by 
\begin{eqnarray}
\kappa = \partial \rho / \partial \mu~,
\label{MF_compressibility}
\end{eqnarray}
with $\rho = \rho_{a}+2\rho_{m}$ the total density.

\subsection{Quantum Monte Carlo simulations}
\label{section3_subB}

The atom-molecule Hamiltonian is simulated by using the stochastic Green function
algorithm \cite{SGF, directedSGF}, an exact QMC
technique that allows canonical  and grand canonical simulations
of the system at zero and finite temperatures, as well as measurements of
many-particle Green functions.  
We treat $L\times L$ lattices with sizes up to $L=14$. 
An inverse temperature of $\beta t=2L$ {\color{black}  allows one to eliminate thermal effects from the QMC results.
We mainly focus on scans at fixed total density  $\rho=\rho_a+2 \rho_m$  in the 
canonical ensemble and calculate the average atomic and molecular densities, 
$\rho_a = \frac{1}{ L^2} \sum_{i} \langle a_i^{\dagger} a_i \rangle~$ and $\rho_m = \frac{1}{ L^2} \sum_{i} \langle m_i^{\dagger} m_i \rangle~$,  respectively, 
and the condensate fraction of atoms and molecules, 
\begin{eqnarray}
\nonumber
C_{a} = \frac{1}{ L^4} \sum_{i,j} \langle a_i^{\dagger} a_j^{\phantom\dagger} \rangle~,\\
C_m = \frac{1}{ L^4} \sum_{i,j} \langle m_i^{\dagger} m_j^{\phantom\dagger} \rangle.
\label{Condensate_fraction}
\end{eqnarray}
The total density $\rho$ is conserved in canonical simulations, but individual
densities $\rho_a$ and $\rho_m$ fluctuate due to the conversion
term Eq.~(\ref{Hamiltonian_atome_molecule}). 
We also calculate the superfluid density 
given by fluctuations of
the winding number \cite{roy}
\begin{equation}
\rho_s = \frac{\langle (W_a + 2W_m)^2\rangle}{4t\beta}
\label{rhos}
\end{equation}

\section{Ground state phase diagrams}
\label{section4}

Without coupling between atoms and molecules, {i.e.} $g=0$, the symmetry of the model is U(1)$\times$U(1) and we expect to observe an atomic (molecular)
Mott insulator for small $t/U$ and integer filling, and atomic (molecular) Bose-Einstein condensate ${\rm BEC_{a}}$ (${\rm BEC_{m}}$) with broken U(1) symmetry for large $t/U$.
Activating the conversion, the symmetry of the model breaks down into a global  U(1)$\times \mathbb{Z}_2$ symmetry corresponding to the transformations 
\begin{eqnarray}
\phi_{i}^m&\to&\phi_{i}^m+\theta \nonumber \\
\phi_{i}^a &\to& \phi_{i}^a+\frac{\theta}{2}+\frac{1}{2}(\sigma+1)\,\pi,  
\end{eqnarray}
with $\sigma=\pm 1$  the Ising variable and $\phi_{i}^a$ and $\phi_{i}^m$ respectively the atomic and 
molecular phases of the fields.  
The U(1) symmetry is a joint one for atomic and molecular phases, and corresponds to total ``mass" conservation with 
density  $\rho= \rho_a+2\rho_m$.
From the mean field point of view,  the average phase of the atoms acquires a finite value in the atomic BEC phase, hence $\langle e^{i\phi_{i}^a} \rangle \neq 0$, 
and consequently $\langle e^{i2\phi_{i}^a} \rangle \neq 0$. 
As a consequence, the molecular phase $\phi_{i}^m$ locked to the nonzero value acquired by the phase of atomic pairs drive the system to 
a joint atomic and molecular BEC (${\rm BEC_{am}}$), and prohibits the existence of an atomic BEC  alone without a molecular condensation.
The reverse is not true: because of the asymmetric nature of the atom-molecule coupling, the molecular condensation does not imply an atomic condensation
and leaves out the $\mathbb{Z}_2$ symmetry. Indeed, the molecular condensation leads to a finite value for the average $\langle e^{i\phi_{i}^m}\rangle$ which
couples to twice the phase of the atoms $\langle e^{i 2 \phi_{i}^a}\rangle \neq 0$ and then fixes the phase of the atoms only modulo $\pi$, {i.e.}  $\phi_{i}^a=\phi_{i}^m/2 \pm \pi$, leading to a fluctuating 
$\phi_{i}^a$ with discrete fluctuations ($\pm \pi$).
Therefore, for large $t/U$, we expect the appearance of two Bose-Einstein condensed phases: a molecular condensate ${\rm BEC_{m}}$ and a mixed atomic-molecular condensate ${\rm BEC_{am}}$.

For small hopping $t/U$, the coupling also strongly affects the Mott insulating phases,  leading to an  atomic-molecular Mott insulating phase (${\rm MI_{am}}$).
The ${\rm MI_{am}}$ phase with $\rho = 2$ is well described by an on-site wave function of the form
 \begin{equation}
|\Psi \rangle = \alpha(\delta/U, g/U) |2,0\rangle + \beta(\delta/U, g/U) |0,1\rangle~,
\label{Wave_function_mixte_insulator}
\end{equation}
in the occupation-number  basis $\{ |n_a, n_m \rangle \}$.
It has been shown that the particle-hole gap $\Delta_{ph}(\delta,g) =  \mu_p(\delta,g) - \mu_h(\delta,g)$, where $\mu_p$ ($\mu_h$) 
is the critical chemical potential to add a particle (hole) to the incompressible phase, is strongly dependent on the conversion parameter $g$.
For a moderate hopping, the most striking feature is that the conversion parameter $g$ can drive the system towards an insulating phase, the
Feshbach insulator (FI), close to Feshbach resonance by opening a particle-hole gap in the ${\rm BEC_{am}}$ phase existing for $g/U \to 0$  \cite{deForges_Roscilde_2015}.
In other words,  the particle-hole gap vanishes in the FI phase when the conversions are suppressed, i.e. $\Delta_{ph}(\delta/U ,g/U \to 0) = 0$, whereas the gap remains finite
in the ${\rm MI_{am}}$ phase when $g=0$.

We first use the MFT described in  Sec.~\ref{section3_subA} for studying the phase diagram and the quantum phase transitions. 
Then, we perform exact QMC simulations described in Sec.~\ref{section3_subB} for a more extensive analysis of the quantum
phase transitions.

\subsection{Mean-field phase diagram}
\label{section4_subA}

The atomic-molecular conversions strongly affect the insulating-BEC transition with two particles per site and give rise to an insulating phase at the tip of the $\rho=2$ Mott lobe.
As a reference, for the standard single species Bose-Hubbard model,  
the Mott-superfluid transition is located at $t_c/U \simeq 0.025$ ($t_c/U \simeq 0.043$) for $\rho=2$ ($\rho=1$),  according to
the mean field method of Sec.~\ref{section3_subA}.
Activating the conversion, Fig.~\ref{MF_insulating_BEC_transition} shows the atomic and molecular condensate fractions $C^{MF}_{a}$ and $C^{MF}_{m}$ 
as functions of the hopping $t/U$ in different regions of the detuning.
\begin{figure}[h!]
\begin{center}
\includegraphics[width=1 \columnwidth]{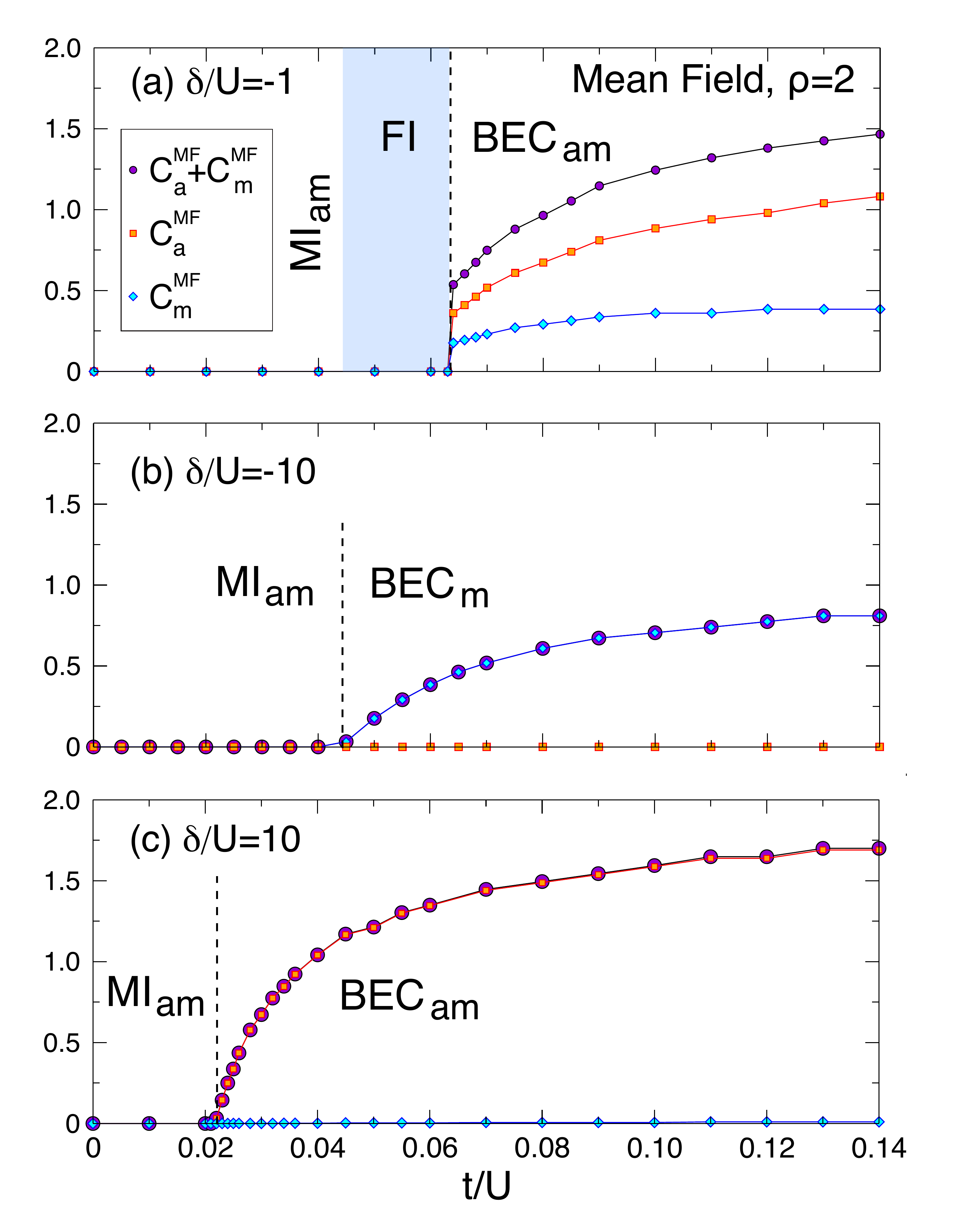}
\caption{(Color online) Mean field insulating-BEC transition for fixed density $\rho=2$ with $g/U=0.8$ close to the resonance (a) $\delta/U=-1$,  (b)  on the molecular side $\delta/U=-10$, and 
(c) on the atomic side  $\delta/U=10$. The FI phase is stabilized by the conversion $g$ whereas the ${\rm MI_{am}}$ phase is  stabilized by the interactions $U$.
The jump in the condensate fractions $C^{MF}_{a}$ and $C^{MF}_{m}$ at  the FI-${\rm BEC_{am}}$ transition indicates a first-order transition, 
associated with the metastable region $t/U \in ] 0.061;0.065 [ $ where the ground state energy exhibits global and local minima.
} 
\label{MF_insulating_BEC_transition}
\end{center}
\end{figure}
For small $t/U$,  the system is in a ${\rm MI_{am}}$ phase for all detuning Fig.~\ref{MF_insulating_BEC_transition}  (a--c) and the particle-hole gap is stabilized by the interaction $U$.
Three scenarios are observed when the hopping  $t/U$ is increased.
Firstly, close to the resonance both $C^{MF}_{a}$ and $C^{MF}_{m}$ turn on simultaneously 
and jump at the transition, indicating the existence of a metastable region at the transition and 
a quantum first-order transition -- see Fig.~\ref{MF_insulating_BEC_transition} (a). 
Clearly, the transition takes place at a critical hopping $t_c/U \simeq 0.062$  bigger than the standard critical hopping of the Mott-superfluid transition without conversions $g=0$ at any
filling. Therefore, the interactions $U$ alone cannot open the particle-hole gap at the tip of the insulating lobe, which is rather stabilized by the conversions $g$: 
the system is in a FI phase. Secondly, only the molecular condensation  $C^{MF}_{m}\neq 0$ occurs on the molecular side -- see Fig.~\ref{MF_insulating_BEC_transition} (b) -- 
the atoms being almost eliminated adiabatically for $\delta/U=-10$. 
Consequently, the transition occurs close to the standard critical value $t_c/U \simeq 0.043$ of
the single species Bose-Hubbard model with $\rho_m \sim 1$.
Lastly, for $\delta/U=10$ (Fig.~\ref{MF_insulating_BEC_transition} (c)), the system is mainly composed by atoms and a mixed condensation occurs when  $t/U$ is increased (the molecular condensate is small but finite).
Although we numerically observe a continuous  ${\rm MI_{am}}$-${\rm BEC_{am}}$ transition,  a weak first-order transition 
is not excluded when fluctuations are taken into account. This, however, does not happen, as we discuss in the following.

We now turn our attention to the phase diagram close to the resonance
with a fixed hopping  $t/U=0.06$  in order to focus on the FI phase.
Figure~\ref{MeanFieldPhaseDiagram}, from Ref.~\cite{deForges_Roscilde_2015}, shows the phase diagram as a function of the detuning $\delta/U$ and of the  chemical potential $\mu/U$.
\begin{figure}
\begin{center}
\includegraphics[width=1 \columnwidth]{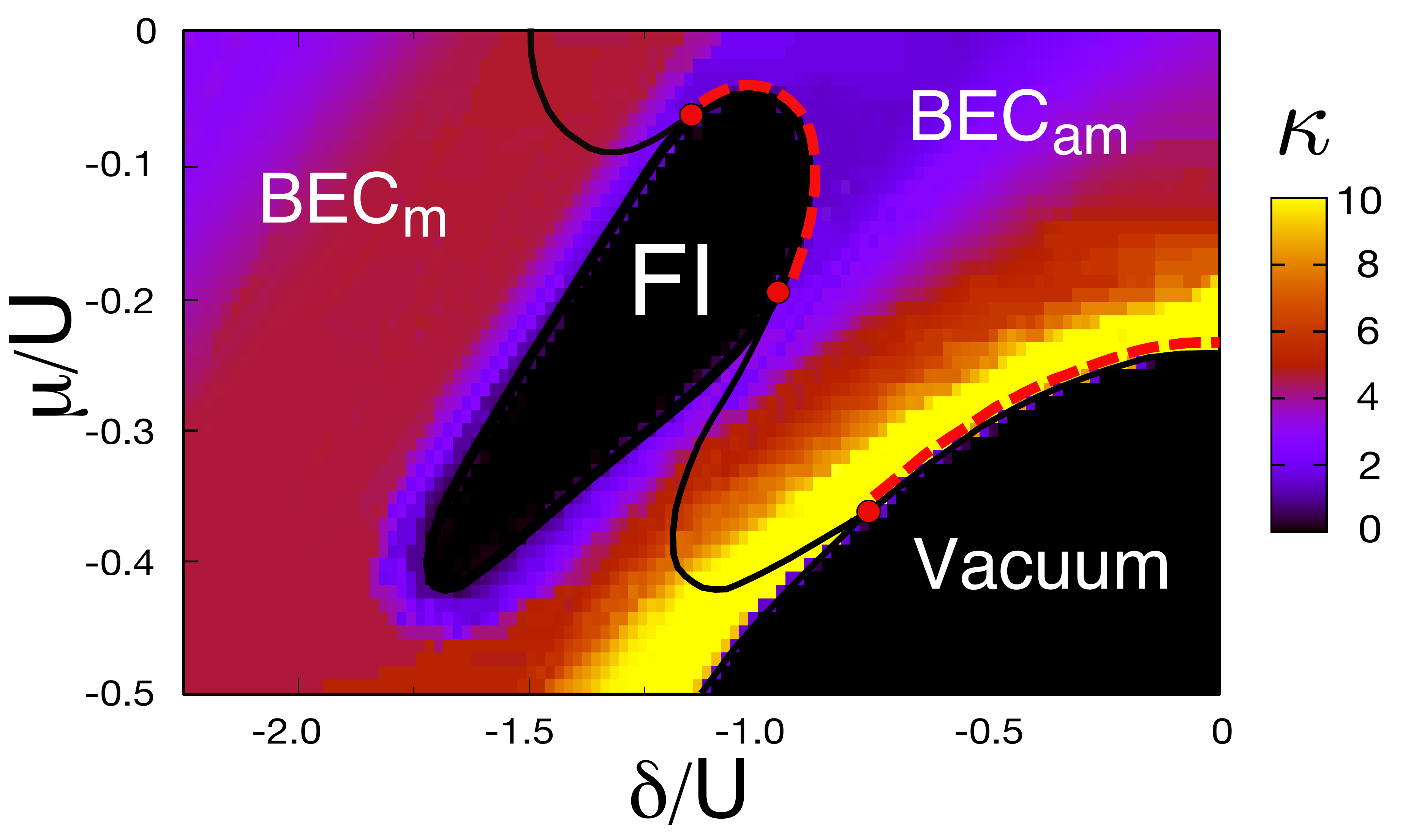}
\caption{(Color online) Mean-field ground state phase diagram, taken from Ref.~\cite{deForges_Roscilde_2015}, with $t/U=0.06$ and $g/U=0.8$ 
(false colors indicate the compressibility $\kappa$ Eq.~(\ref{MF_compressibility})). 
The following phases appear in the phase diagrams: Feshbach insulator (FI) with $\rho=2$,  molecular  (${\rm BEC_{m}}$) and atomic-molecular  (${\rm BEC_{am}}$) condensate.  
Second-order transitions are denoted by solid  black lines, red dashed lines indicate first-order transitions, and red dots denote tricritical points. 
}
\label{MeanFieldPhaseDiagram}
\end{center}
\end{figure}
The incompressible region (black region in Fig.~\ref{MeanFieldPhaseDiagram})  reveals the existence of the particle-hole gap of the FI phase with $\rho=2$.
The molecular condensation ${\rm BEC_{m}}$ and the mixed condensation ${\rm BEC_{am}}$ are also observed in the phase diagram.
First-order transitions, indicated by red dashed lines in Fig.~\ref{MeanFieldPhaseDiagram}, are systematically observed when both atomic and molecular 
order parameters are simultaneously turned on, {i.e.} when the global symmetry of the model U(1)$\times \mathbb{Z}_2$ is destroyed \cite{footnote:U1Z2}.
The first-order nature of the transition is not specific to the transition to the FI, but it appears to be generic for all insulating-${\rm BEC_{am}}$ transitions \cite{Radzihovskyetal2008}.
Although this phase diagram was discussed in Ref.~\cite{deForges_Roscilde_2015}, the phase transitions have not yet been properly examined using an exact method.

Changing the detuning $\delta/U$,  {i.e.} the control parameter in the experiment, can drive the system into different phases, leading to quantum phase transitions without changing 
the hopping parameter $t/U$. This gives the experimental possibility to observe multiple transitions and more specifically the first-order FI-${\rm BEC_{am}}$ transition.
The direct observation of the density profile for a fix detuning would  give rise to intriguing shapes  since many first-order transitions are involved.
According to the local density approximation, the density profile is obtained by a vertical scan of the phase diagram by changing the chemical potential $\mu/U$. 
Figure~\ref{Vertical_cut_mean_field_phase_diagram} shows the condensates and the densities for such a vertical cut with $\delta/U=-1.0$.
\begin{figure}
\begin{center}
\includegraphics[width=1 \columnwidth]{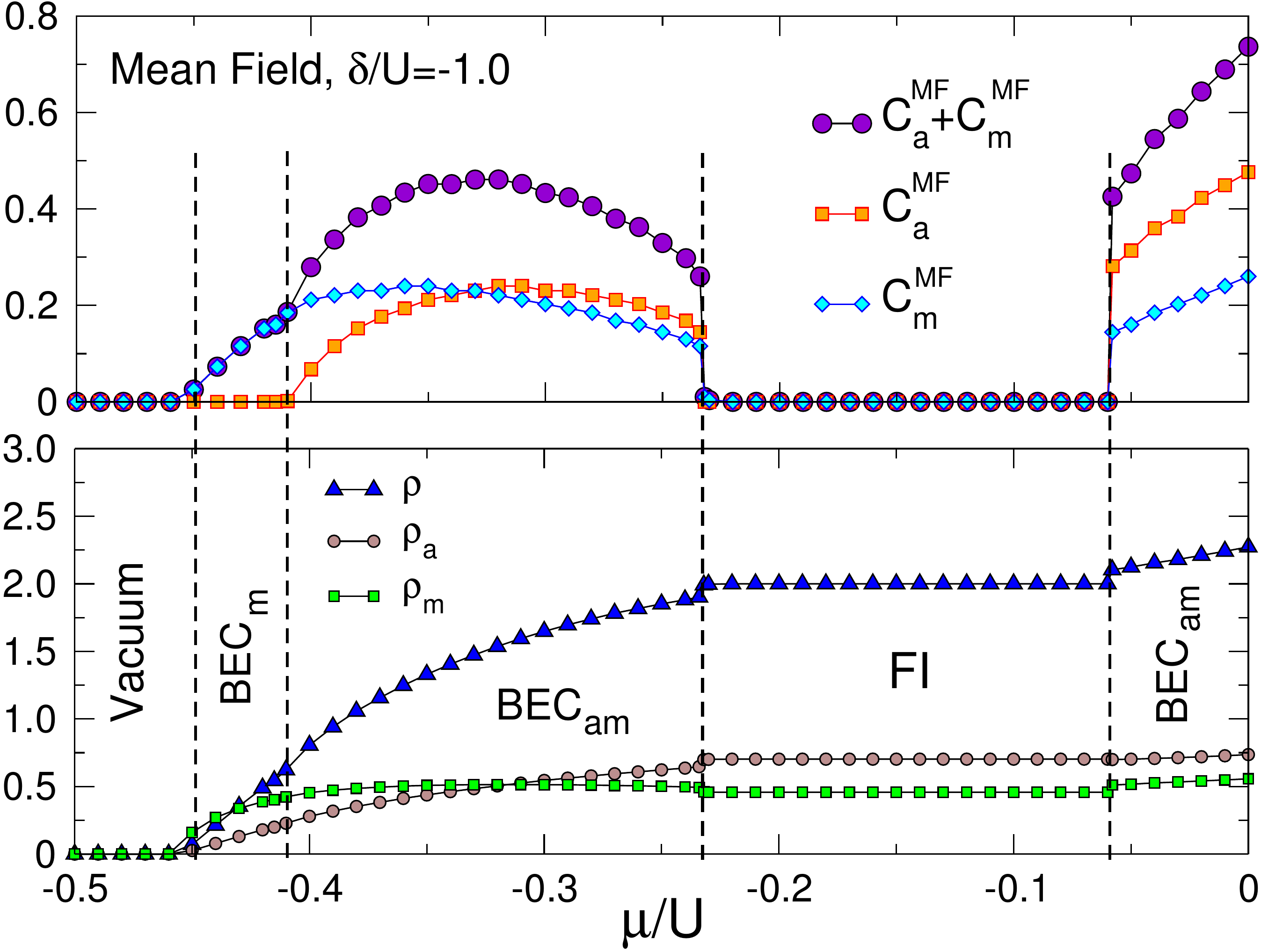}
\caption{(Color online)  Vertical cut of the mean-field phase diagram in Fig.~\ref{MeanFieldPhaseDiagram} for $\delta/U=-1.0$.
The  ${\rm BEC_{am}}$-FI transition is first-order whereas the other transitions are continuous.}
\label{Vertical_cut_mean_field_phase_diagram}
\end{center}
\end{figure}
The system evolves continuously from vacuum to ${\rm BEC_{m}}$ and to ${\rm BEC_{am}}$ when $\mu/U$ increases, and 
all the quantities jump at the first-order ${\rm BEC_{am}}$-FI transition. Note that both atomic and molecular densities, $\rho_a$ and $\rho_m$, 
reach a noninteger plateau in the FI phase, whereas the total density is  integer $\rho=\rho_a+2\rho_m=2$.

The mean field analysis reports a rich physics attributed to the conversion term in Eq.~(\ref{Hamiltonian_atome_molecule}) 
but does not allow the classification of the transitions which requires the calculation of the correlation functions.

\subsection{Quantum Monte Carlo simulations}
\label{section4_subB}

The MFT results are qualitatively confirmed by our QMC simulations. 
\begin{figure}
\begin{center}
\includegraphics[width=1 \columnwidth]{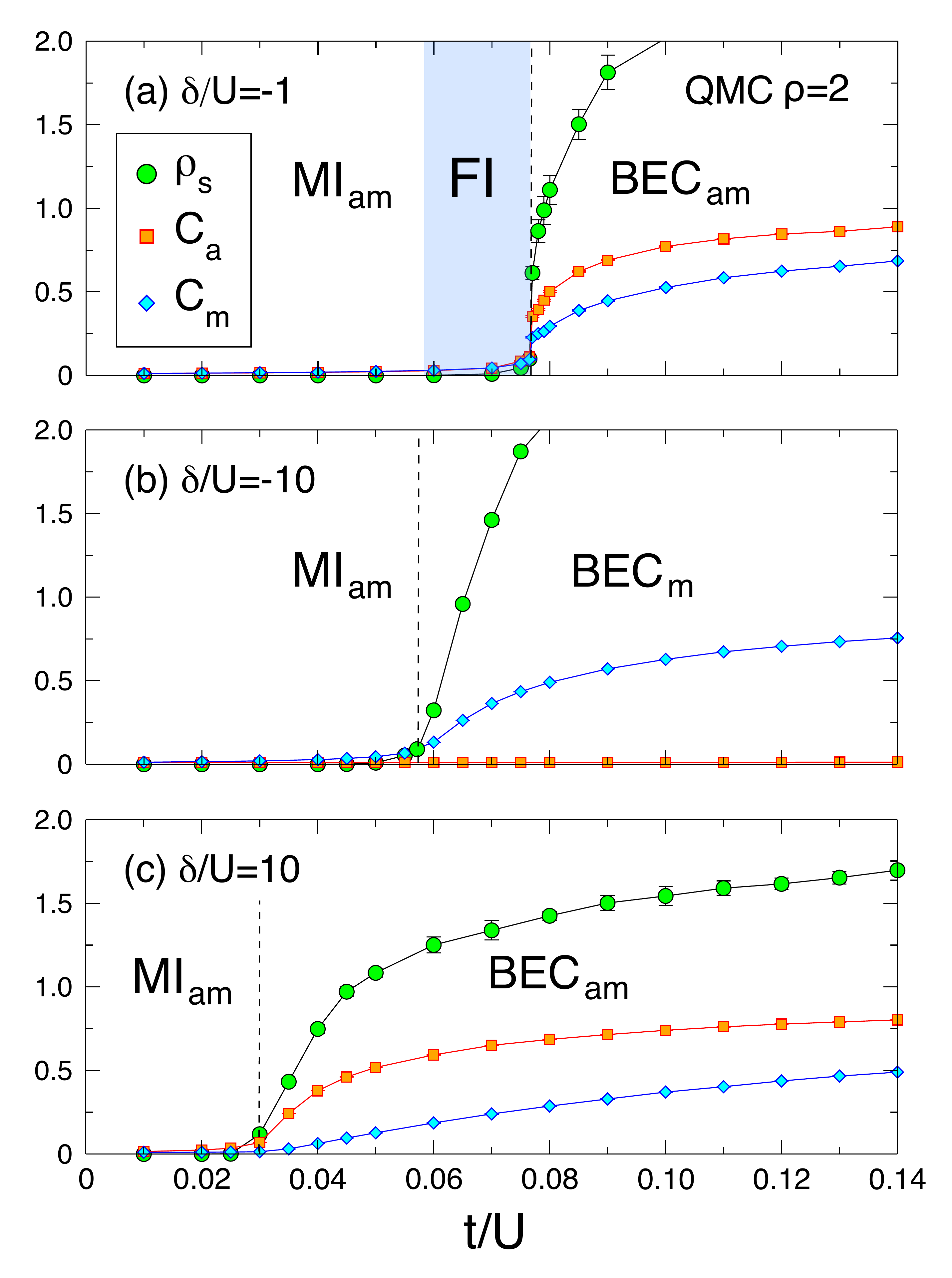}
\caption{(Color online) QMC simulations of the insulating-BEC transition with $L=10$ for fixed density $\rho=2$ and $g/U=0.6$, $\beta t=20$ 
close to the resonance (a) $\delta/U=-1$,  (b)  on the molecular side $\delta/U=-10$, and 
(c) on the atomic side  $\delta/U=10$. These results are in good qualitative agreement with the mean field predictions of Fig.~\ref{MF_insulating_BEC_transition}.
(a) Close to the resonance, the conversion stabilizes the FI phase. The discontinuities in the atomic and molecular condensates,  $C_a$ and $C_m$ (Eq.~\ref{Condensate_fraction}), and 
in the superfluid density $\rho_s$ (Eq.~\ref{rhos}), indicate a first-order transition.
} \label{QMC_insulating_BEC_transition}
\end{center}
\end{figure}
Figure~\ref{QMC_insulating_BEC_transition} (a) shows that, upon lowering $t/U$, the atomic and molecular condensates exhibit a clear jump at the tip of the FI, witnessing the first-order 
nature of the FI-${\rm BEC_{am}}$  transition. The jump is also observed in the superfluid density. The transition occurs  
at a critical $t_c/U \simeq 0.076$ value well above the standard critical hopping of the MI-SF transition without conversions $g=0$ at any filling 
(e.g. $t_c/U \simeq 0.059$ for the single-species Bose-Hubbard model at the MI-SF transition with $\rho=1$ \cite{Capogrossoetal2008}), then proving the crucial  contribution 
of the conversions in the particle-hole gap stabilization. 
The ${\rm MI_{am}}$-FI crossover was investigated in Ref.~\cite{deForges_Roscilde_2015}.
Far on the molecular side -- Fig.~\ref{QMC_insulating_BEC_transition} (b) -- only the molecular condensation occurs, and the continuous ${\rm MI_{am}}$-${\rm BEC_{m}}$ transition 
takes place close to $t_c/U\simeq 0.054$. Finally, on the atomic side -- Fig.~\ref{QMC_insulating_BEC_transition} (c) -- we do not observe a 
jump at the ${\rm MI_{am}}$-${\rm BEC_{am}}$ transition at $t_c/U\simeq 0.03$.

To confirm the presence of a first-order quantum phase transition near
the tip of the FI lobe, we perform a finite size analysis of the condensates $C_a$, $C_m$ and of the 
superfluid density $\rho_s$. Indeed, since the correlation length does not diverge at a first-order transition, the jump should increase with the 
system size for small systems, and then saturate for big sizes.
\begin{figure}
\begin{center}
\includegraphics[width=1 \columnwidth]{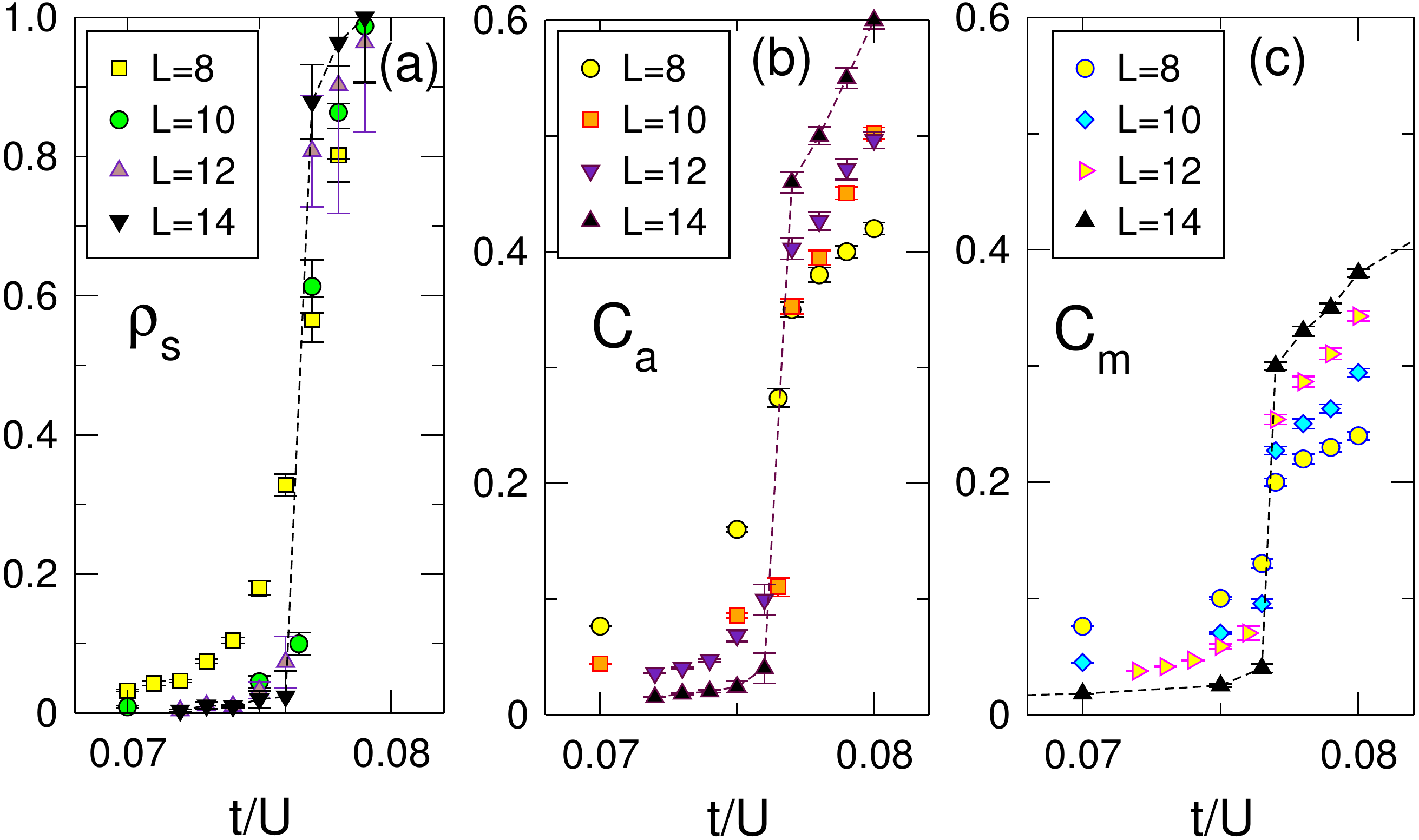}
\caption{(Color online) Finite size analysis of the (a) superfluid density $\rho_s$ and of the condensates (b) $C_a$ and (c) $C_m$ 
at the FI-${\rm BEC_{am}}$  transition using QMC simulations with $\delta/U=-1$, $g/U=0.6$, $\beta t=2L$ and $\rho=2$
for  linear sizes $L=8, 10, 12, 14$.
The jump at the transition increases with $L$.} 
\label{Finite_size_scaling_FI_BECam_transition}
\end{center}
\end{figure}
Figure~\ref{Finite_size_scaling_FI_BECam_transition} clearly shows that the jump increases with the linear system size $L$, 
indicating a first-order  FI-${\rm BEC_{am}}$  phase transition.
This conclusion is strengthened by  QMC grand canonical simulations, see Fig.~\ref{rho_vs_mu_GC_L8_tsurU0p074_gsurU0p6}, 
where the  density jumps at the FI-${\rm BEC_{am}}$ transition indicating a metastable region and a first-order transition at the tip of the FI lobe.
\begin{figure}
\begin{center}
\includegraphics[width=1 \columnwidth]{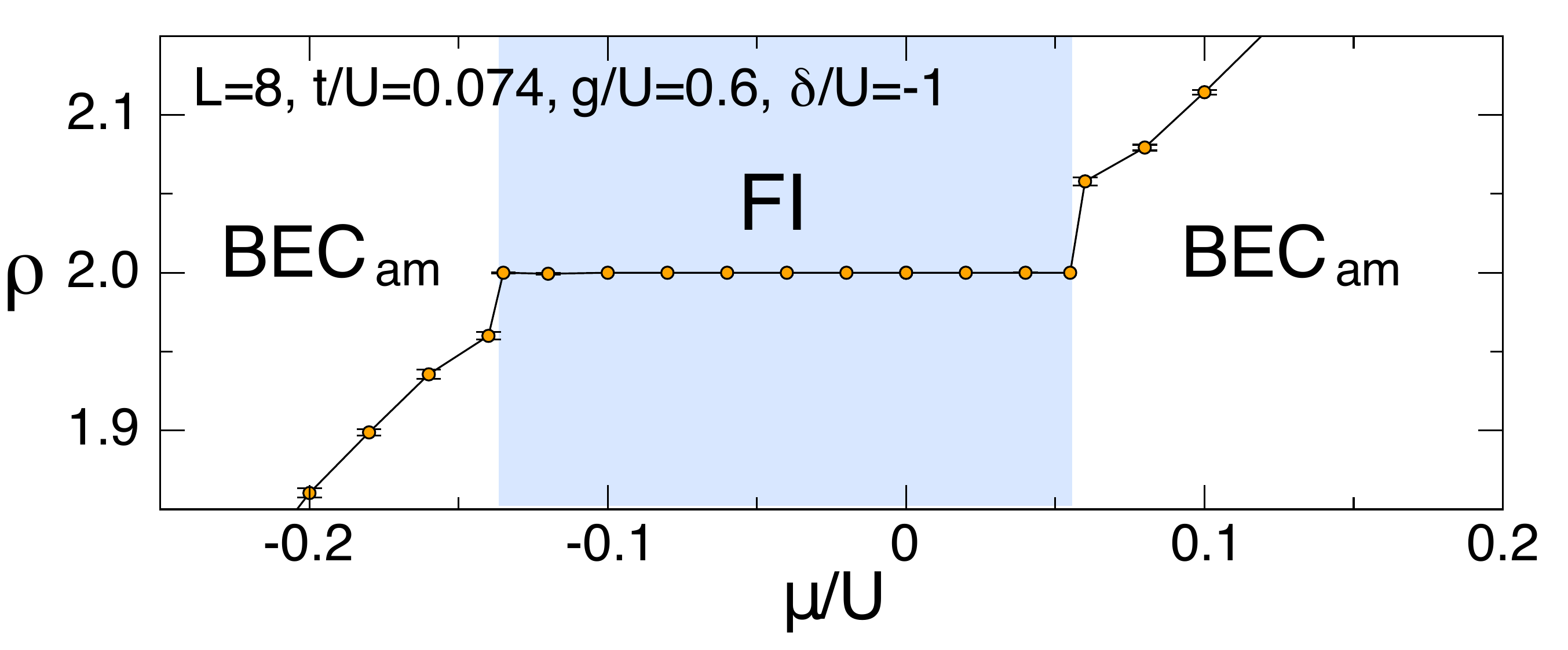}
\caption{(Color online) QMC grand canonical simulations for $L=8$ and the same other parameters as in Fig.~\ref{Finite_size_scaling_FI_BECam_transition}.
The density $\rho$ is discontinuous at the FI-${\rm BEC_{am}}$ transition. 
} 
\label{rho_vs_mu_GC_L8_tsurU0p074_gsurU0p6}
\end{center}
\end{figure}

We now investigate the quantum phase transitions keeping both  hopping and conversion fixed and varying the detuning $\delta/U$.
Since the FI phase is stabilized for even densities only, two different behaviors are observed for even and odd densities.
\begin{figure}
\begin{center}
\includegraphics[width=1 \columnwidth]{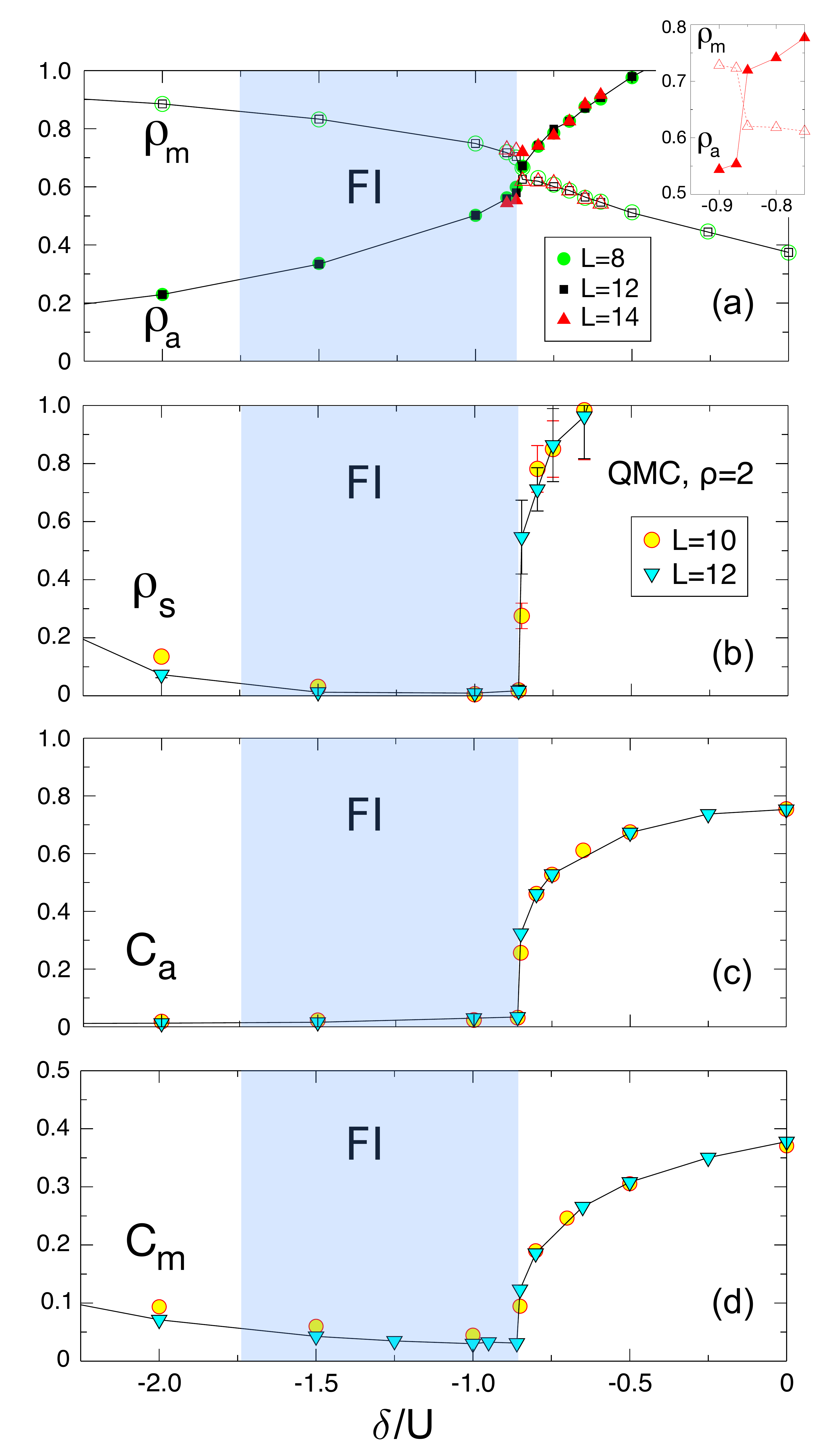}
\caption{(Color online) QMC canonical simulations at fixed total density  $\rho=2$,  with  $g/U=0.6$, $\beta t=2L$ and $t/U=0.07$.
(a) Atomic $\rho_a$ and molecular $\rho_m$ densities as  functions of detuning: both densities jump at the first-order FI-${\rm BEC_{am}}$ transition (see zoom in inset for $L=14$); 
A jump is also observed  in the (b) superfluid density $\rho_s$,  as well as in the (c) atomic $C_a$ and (d) molecular $C_m$ condensates when the size of the system increases.
} 
\label{QMC_densities_rhos_condensates_vs_delta}
\end{center}
\end{figure}
We first discuss the case with two particles per site.
Starting in the ${\rm BEC_{m}}$ phase for large negative $\delta/U$, the system evolves firstly in the FI phase, and then in the  ${\rm BEC_{am}}$ phase when 
increasing the detuning -- see Fig.~\ref{QMC_densities_rhos_condensates_vs_delta}.
As expected, the atomic density $\rho_a$ increases with the detuning, see Fig.~\ref{QMC_densities_rhos_condensates_vs_delta} (a) and
both atomic and molecular densities jump at the first-order FI-${\rm BEC_{am}}$ transition, see inset Fig.~\ref{QMC_densities_rhos_condensates_vs_delta} (a). 
This jump is also clearly observed in the  superfluid density and in the condensate fractions at the FI-${\rm BEC_{am}}$ transition 
(Fig.~\ref{QMC_densities_rhos_condensates_vs_delta} (b-d)).

As shown in Fig.~\ref{QMC_densities_rhos_condensates_vs_delta_rho1}, the FI phase cannot be stabilized for $\rho=1$ 
and a ${\rm BEC_{m}}$-${\rm BEC_{am}}$ transition is induced upon increasing the detuning $\delta/U$. 
\begin{figure}
\begin{center}
\includegraphics[width=1 \columnwidth]{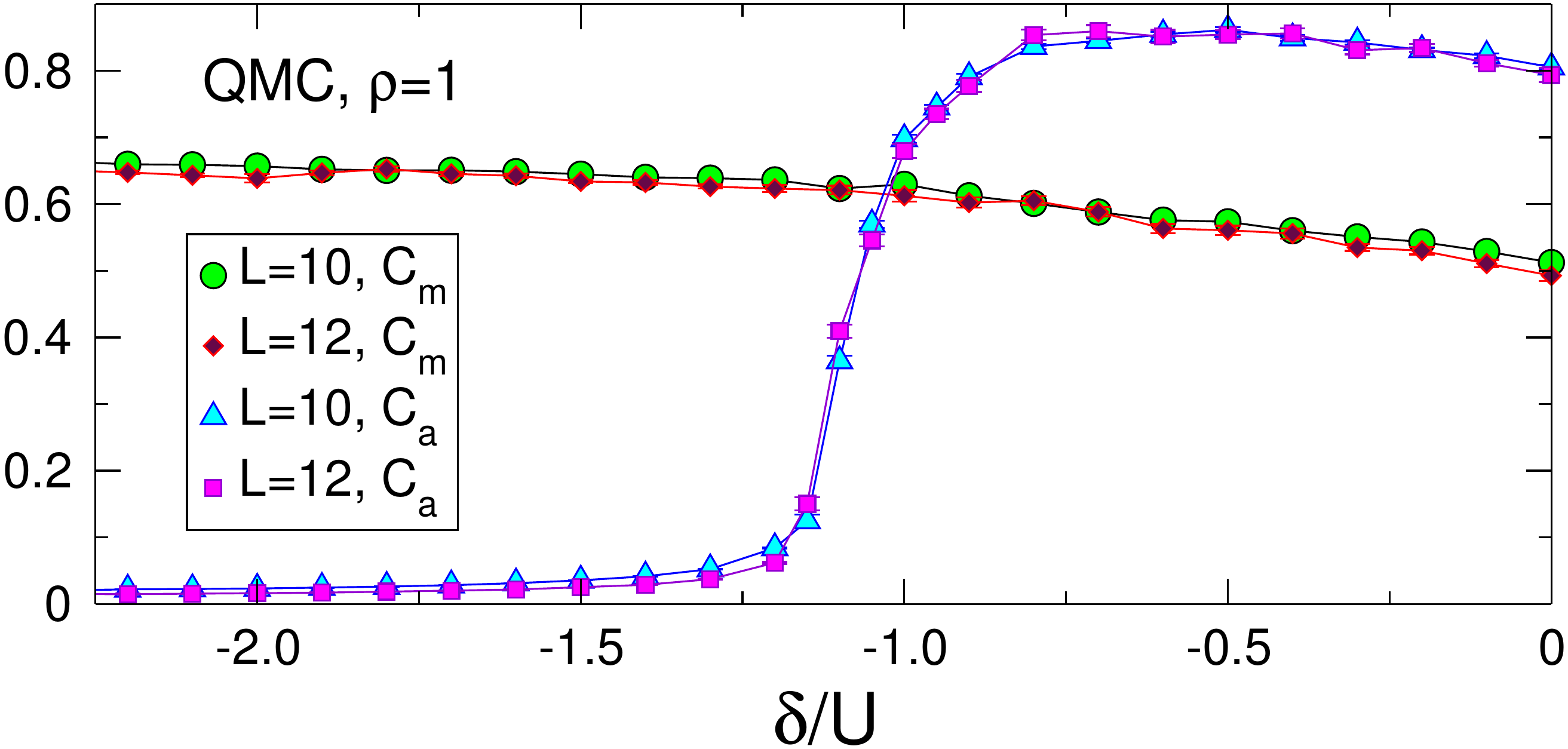}
\caption{(Color online) QMC canonical simulations of the atomic and molecular condensates as
functions of the detuning $\delta/U$ for $\rho=1$, $g/U=0.6$,  $t/U=0.07$, and $\beta t=2L$ and two linear sizes $L=10$
and $L=12$. The system evolves from a ${\rm BEC_{m}}$ to a ${\rm BEC_{am}}$ phase when increasing the  detuning $\delta/U$.
} 
\label{QMC_densities_rhos_condensates_vs_delta_rho1}
\end{center}
\end{figure}
This transition, captured also at the mean-field level, is related to the breaking of the $\mathbb{Z}_2$ symmetry associated with the 
phase of the atomic field, and it is therefore expected to belong to the 3D Ising universality class. 
While the universality class cannot be correctly reproduced at the mean-field level, our QMC simulations show a very convincing 
scaling of the condensate fraction as $C_a = L^{-2\beta/\nu} f(L^{1/\nu}|\delta-\delta_c|/U)$ with exponents $\beta$ and $\nu$ 
belonging to the 3D Ising universality class -- see Fig.~\ref{Scaling_condensates_transitionBKT_Ising} (b) for $\rho=1$. 
The ${\rm BEC_{m}}$-${\rm BEC_{am}}$ transition also belongs to the 3D Ising universality class for $\rho \neq1$
according to our QMC simulations -- e.g. see Fig.~\ref{Scaling_condensates_transitionBKT_Ising} (a) for $\rho=0.5$. 
\begin{figure}
\begin{center}
\includegraphics[width=1 \columnwidth]{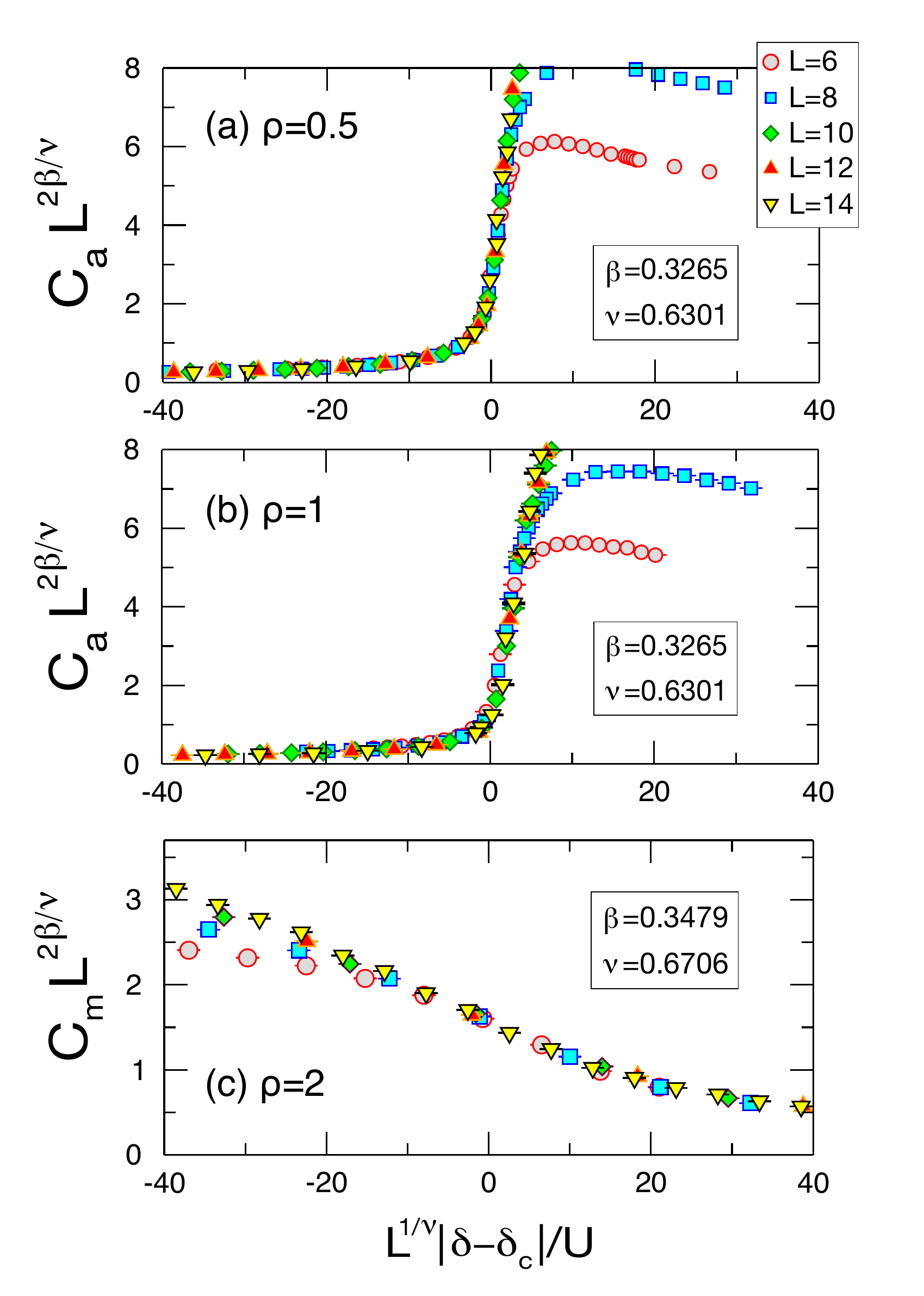}
\caption{(Color online) Scaling plots of the atomic and molecular  condensates for the quantum phase 
transition from $\rm BEC_m$ to $\rm BEC_{am}$ with  (a)  $\rho=0.5$, (b) $\rho=1$, and (c)  from $\rm BEC_m$ to  FI with  $\rho=2$, 
as obtained via QMC simulations with $\beta t=2L$. 
Critical exponents of the (a-b) 3D Ising  and (c) 3D $XY$  universality classes \cite{PelissettoV2002}, 
cited in the boxes, are used for the scaling.} 
\label{Scaling_condensates_transitionBKT_Ising}
\end{center}
\end{figure}
Similarly to the single-species Bose-Hubbard model,  the scaling of $C_m$ is found to be consistent with the 3D $XY$ universality class 
at the $\rho=2$ $\rm BEC_m$-FI transition where only the U(1) symmetry is spontaneously restored -- see Fig.~\ref{Scaling_condensates_transitionBKT_Ising} (b). 
The other transitions in the phase diagram of Fig.~\ref{MeanFieldPhaseDiagram}, {i.e.} vacuum-$\rm BEC_m$ and FI-$\rm BEC_m$  with $\rho \neq 2$,
are second order (not shown). The order and the universality class of the quantum phase transitions of the phase diagram in
Fig.~\ref{MeanFieldPhaseDiagram} are summarized in Table~\ref{tab1}.

\begin{table}
\begin{tabular}{c c c}
\hline
\hline
Quantum Phase Transition & Order & Universality Class \\
\hline
$\rm BEC_m$-$\rm BEC_{am}$ & $2^{\rm nd}$& 3D Ising\\
$\rm BEC_m$-FI $\rho=2$ & $2^{\rm nd}$& 3D XY\\
$\rm BEC_m$-FI $\rho \neq 2$ & $2^{\rm nd}$& Mean field \\
Vacuum-$\rm BEC_{m}$ & $2^{\rm nd}$ & Mean field  \\
Vacuum-$\rm BEC_{am}$ & $1^{\rm st}$& $\O$ \\
FI-$\rm BEC_{am}$ & $1^{\rm st} $& $\O$ \\
\hline
\hline
\end{tabular}
\caption{Order and universality class of the quantum phase transitions of 
the phase diagram Fig.~\ref{MeanFieldPhaseDiagram}.  \label{tab1}}
\end{table}

\section{Thermal phase diagram}
\label{section5}

In two dimensions,, the  Bose-Einstein condensation at finite
temperature is impossible in the proper sense \cite{Mermin}, leaving
space to quasicondensation via a BKT transition \cite{Josebook}, associated with the unbinding
transition of pairs of topological excitations (vortices and antivortices). 
In this context, the coherent coupling
between atoms and molecules establishes a correlation among the topological
defects in the phase patterns of both species which brings interesting features \cite{deForges2016}.
We first analyze the thermal phase diagram for $\rho=2$, and then turn to an analysis of the thermal phase transitions.

\subsection{Thermal phase diagram for $\rho=2$}
\label{section5_subA}

The ground state analysis  (Sec.~\ref{section4})  revealed the possibility to stabilize an insulating phase, the FI phase, with a finite particle-hole gap opened  
by the conversions between atoms and molecules for even total density.  The FI phase evolves either in a $\rm BEC_m$ phase when decreasing the 
detuning $\delta$, or in a  mixed  $\rm BEC_{am}$ phase when increasing $\delta$ (see Fig.~\ref{QMC_densities_rhos_condensates_vs_delta}).  
The thermal phase diagram, plotted in  Fig.~\ref{Thermal_Phase_Diagram_rho2}, shows the evolution of the phases when activating the thermal effects.
\begin{figure}
\begin{center}
\includegraphics[width=1 \columnwidth]{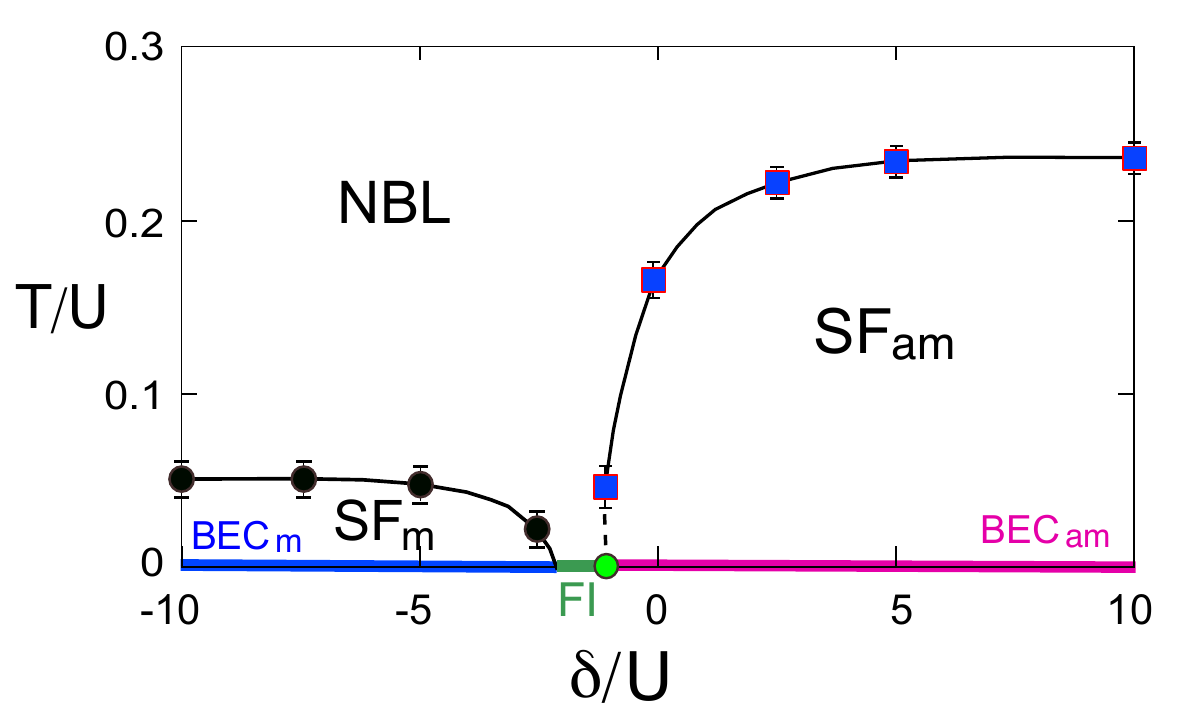}
\caption{(Color online)  QMC thermal phase diagram with $\rho=2$,  $g/U=0.6$ and $t/U=0.07$: the BEC phases existing strictly 
at $T=0$ become superfluid (SF) and then normal Bose liquid (NBL) when heating the system, keeping the detuning constant,
whereas the FI phase crossed over the NBL phase. 
Black circles and blue squares are obtained using finite size scaling analysis, 
see respectively Figs.~\ref{Scaling_molecular_condensate_and_rhos_vs_Temperature} and \ref{Scaling_atomique_molecular_condensate_vs_Temperature}.
The green point indicates a quantum first-order transition;  the dashed line indicates a possible first-order transition.}
\label{Thermal_Phase_Diagram_rho2}
\end{center}
\end{figure}
As expected, the molecular (mixed) condensate become molecular (mixed) superfluid at low temperature with $\rho_s \neq 0$  
and  the system is in a normal Bose liquid (NBL) at high temperature for all detuning $\delta$. 
The critical temperature at the SF-NBL transition is determined using finite size analysis, see Sec.~\ref{section5_subB}.

Note that  the mixed superfluid $\rm SF_{am}$ and the molecular superfluid $\rm SF_{m}$, separated by the FI phase at $T=0$, remain 
well separated for all temperature and  the insulating FI phase crossed over the NBL phase when the temperature is increased.
Interestingly, although $t_a=t_m=t$,  the $\rm SF_{am}$ phase is more robust with respect to the thermal effects, compared to the single component $\rm SF_{m}$ superfluid.
This can be qualitatively explained  by looking at the atomic (molecular) characteristic interacting scales $J_{a(m)} \sim  \rho_{a (m)} t$, 
where $\rho_a =2$ far on the atomic side ($\delta \to +\infty$) and $\rho_m =1$ far on the molecular side ($\delta \to -\infty$). 
Therefore, the interacting scale in the  $\rm SF_{am}$  phase is twice as large as the one in the  $\rm SF_{m}$ phase.
The same behavior has been observed at the mean field level in 3D \cite{Radzihovskyetal2004}.

\subsection{Quantum-to-classical first-order phase transition and nonconventional BKT transitions}
\label{section5_subB}

The quantum first-order transition between the FI and the mixed condensate $\rm BEC_{am}$ requires a specific attention since 
it is not excluded that the metastability region persists at $T \neq 0$. Indeed, the discontinuity  in the superfluid density at the disordered-$\rm SF_{am}$  transition -- 
discontinuity associated with the existence of the metastability  region --
persists at finite temperature, see Fig.~\ref{jump_in_rhos_vs_Temperature} (a).
\begin{figure}[!h]
\begin{center}
\includegraphics[width=1 \columnwidth]{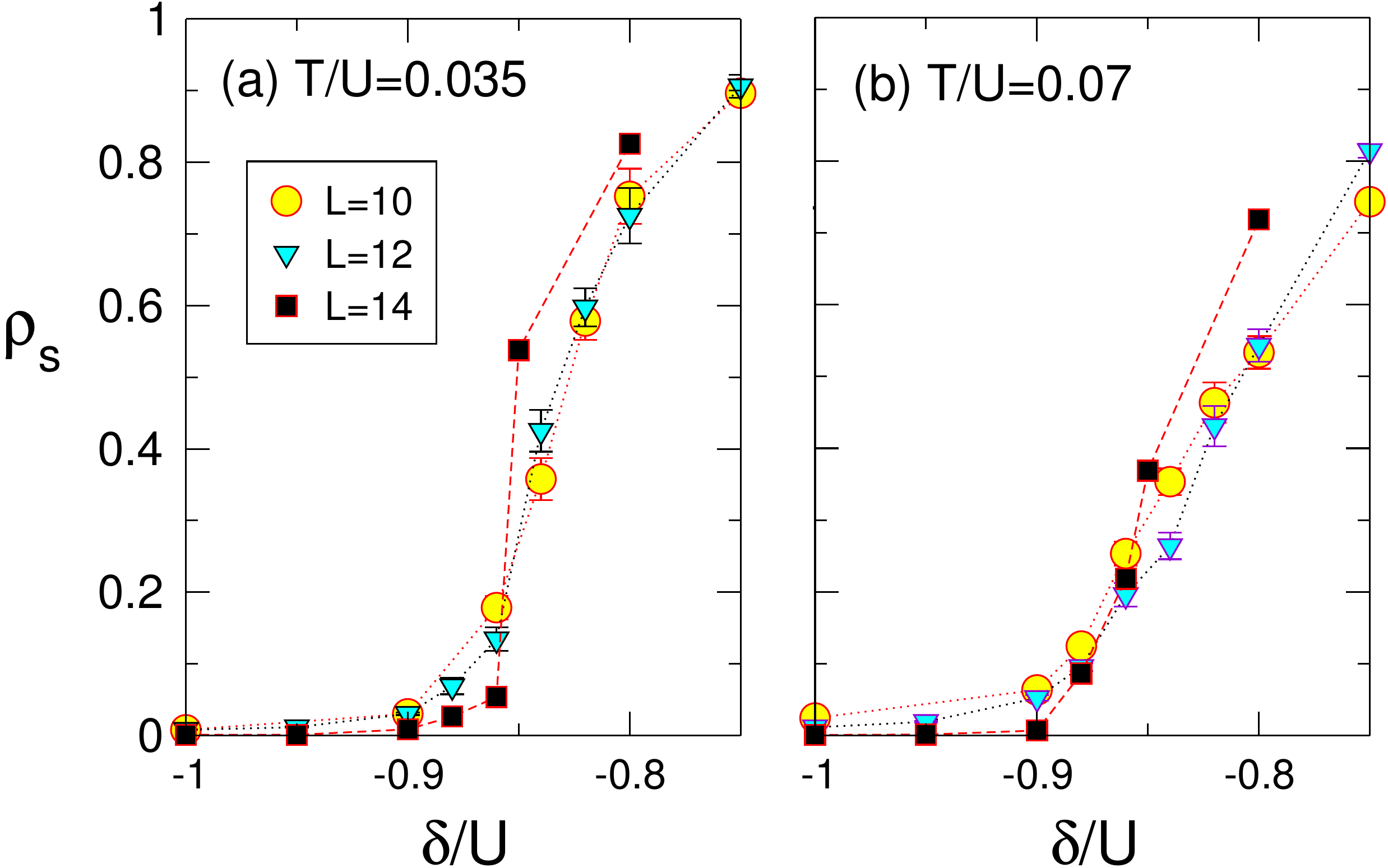}
\caption{(Color online) Superfluid density as a function of the detuning at the $\rho=2$ disordered-$\rm SF_{am}$ transition 
for finite temperatures with $L=10, 12 , 14$,   $g/U=0.6$ and $t/U=0.07$. The discontinuity in $\rho_s$, observed for (a) $T/U=0.035$, decreases with the temperature.
No jump is observed for (b)  $T/U=0.07$.} 
\label{jump_in_rhos_vs_Temperature}
\end{center}
\end{figure}
Interestingly enough, the discontinuity in $\rho_s$ observed at $T/U=0.035$ in Fig.~\ref{jump_in_rhos_vs_Temperature} (a) -- a temperature close to the critical
temperature of the $\rm SF_{m}$-NBL transition far on the molecular side, reinforces the idea that the quantum first-order FI-$\rm BEC_{am}$  transition 
becomes a classical first-order NBL-$\rm SF_{am}$ transition. 
As previously discussed,  the discontinuity in $\rho_s$ increases with the system size at a 
first-order transition. This behavior is clearly observed for $T/U=0.035$ in Fig.~\ref{jump_in_rhos_vs_Temperature} (a).
However, we do not observe other clear signals of a first-order transition (two-peak structures in histograms or negative compressibility) 
and therefore we are not able to discern whether the  phase transition indeed is first order or not.

We now concentrate on vertical slices of the phase diagram in Fig.~\ref{Thermal_Phase_Diagram_rho2}, starting in the condensed phase at low temperature,
keeping the detuning fixed and  varying the temperature. 
It is well known that the loss of the  quasi-long-range coherence at a BKT transition is associated with the unbinding of vortices and antivortices and with 
a scaling of the  quasicondensate such that $C \propto L^{-\eta}$ with the critical exponent $\eta(T_{BKT})=1/4$ \cite{lebellac}.
Furthermore, the superfluid density satisfies the universal jump $\rho_s /T_{\rm BKT}= 2/\pi$ at the transition \cite{Nelson_Kosterlitz_1977}.
To avoid any confusion, we stress that the universal jump in $\rho_s$ at a BKT transition is only observed in the thermodynamic limit 
(e.g., see Fig.~\ref{Scaling_molecular_condensate_and_rhos_vs_Temperature} (b)), and therefore 
cannot be confused with the discontinuity at a first-order transition observed for $L \sim 10$.
In the case of coupled fields, the BKT transition implies a more complex mechanism since the topological defects are coupled, 
as for e.g.,  unbinding of half-vortices instead of the usual  integer vortices \cite{Mukerjee_2006, Wessel_2011, Yang_2011, deForges2016}.

We first discuss the BKT transition far on the molecular side, with $\delta/U=-10$ -- see Fig.~\ref{Scaling_molecular_condensate_and_rhos_vs_Temperature}.
\begin{figure}
\begin{center}
\includegraphics[width=1 \columnwidth]{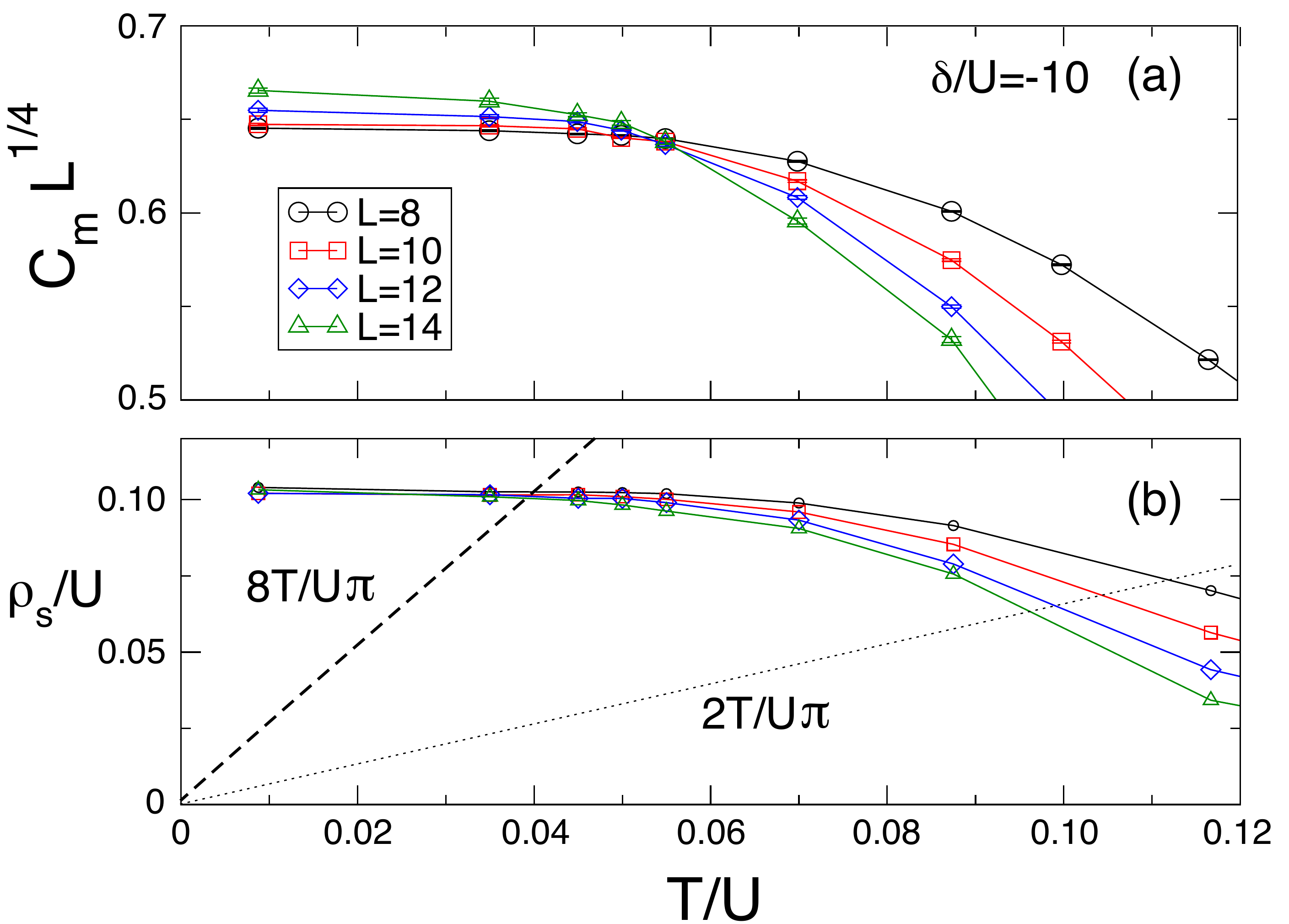}
\caption{(Color online) (a) Scaling of the molecular condensate $C_m$ and (b) superfluid density as functions of temperature 
for different sizes and $\rho=2$, $g/U=0.6$, $t/U=0.07$, and $\delta/U=-10$. 
The critical temperature extracted from scaling analysis with the BKT exponent in (a) is consistent with an anomalous $8T_{BKT}/\pi$  stiffness jump instead of $2 T_{BKT}/\pi$.
} 
\label{Scaling_molecular_condensate_and_rhos_vs_Temperature}
\end{center}
\end{figure}
The critical temperature $T_{BKT}/U \sim 0.045$, in agreement with Ref.~\cite{Capogrossoetal2008}, is determined from the finite size 
scaling of the molecular condensate fraction $C_m(T_{BKT}) \propto L^{-1/4}$ with system sizes up to $L=14$ -- see Fig.~\ref{Scaling_molecular_condensate_and_rhos_vs_Temperature} (a). 
Moreover, the BKT transition is found to be consistent with an
anomalous $8T_{BKT}/\pi$  stiffness jump at $T_{BKT}$ instead of $2 T_{BKT}/\pi$ -- see 
Fig.~\ref{Scaling_molecular_condensate_and_rhos_vs_Temperature} (b).
This anomalous jump is easily understood by rewriting the superfluid density Eq.~(\ref{rhos}) with a vanishing atomic winding number $\langle W_a \rangle=0$ on the molecular side, 
leading to $\rho_s = \frac{\langle (2W_m)^2\rangle}{4t\beta} = 4 \rho_{s0}$, with $\rho_{s0}$ the single component winding number.
That immediately gives the factor 4 involved in the anomalous stiffness jump observed.

On the atomic side, the situation is more complex since both atomic and molecular fields are coupled in a regime
of a quasi-long range order in their correlation.
For large positive detuning, we expect the atomic condensate fraction $C_a$ to satisfy the standard BKT scaling.
This behavior is depicted for $\delta/U=10$  in Fig.~\ref{Scaling_atomique_molecular_condensate_vs_Temperature} (a), where
$C_a(T_{BKT}) \propto L^{-1/4}$. 
\begin{figure}
\begin{center}
\includegraphics[width=1 \columnwidth]{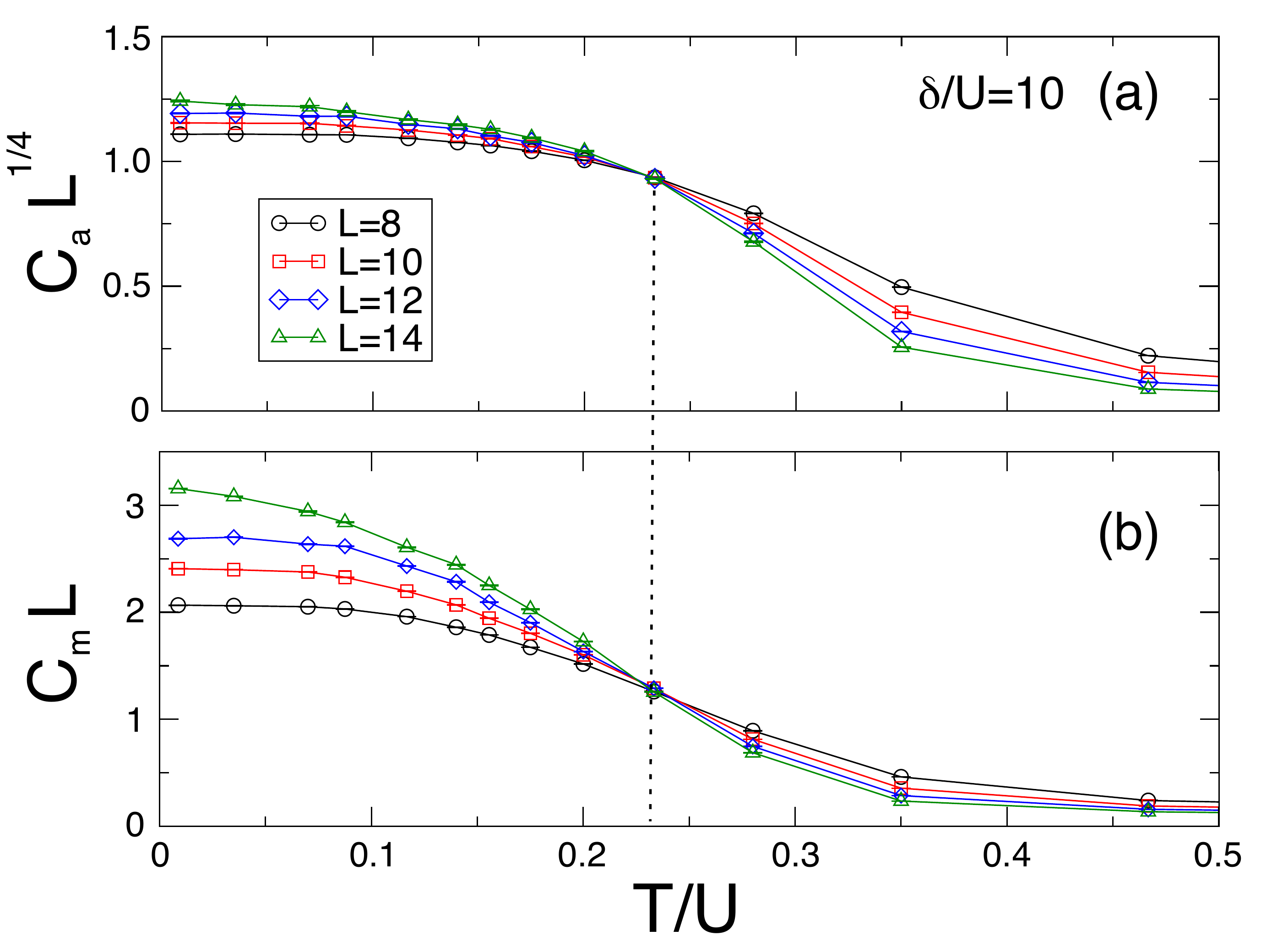}
\caption{(Color online) Scaling of the (a) atomic and (b) molecular condensate as functions of temperature 
for different sizes with $\rho=2$, $g/U=0.6$, $t/U=0.07$, and $\delta/U=10$.}  
\label{Scaling_atomique_molecular_condensate_vs_Temperature}
\end{center}
\end{figure}
This transition is in agreement with the standard $2 T_{BKT}/\pi$ universal stiffness jump (not shown).
As discussed in Sec.~\ref{section4} for $T=0$, the molecular phase $\phi_{\bm r}^m$ is locked (from the mean field point of view) to the nonzero value acquired by the phase of 
atomic pairs $\langle e^{i2\phi_{\bm r}^a} \rangle$, since  the average phase of the atoms acquires a finite value $\langle e^{i\phi_{\bm r}^a} \rangle \neq 0$.
As a consequence, the topological defects in the atomic and molecular field are coupled also at finite temperature, 
leading to the appearance of vortices in the molecular field due to the conversion term \cite{deForges2016}.
For large coupling, the molecular field is expected to quasicondense at the atomic BKT transition in the same way as atom pairs condense.
Therefore, the molecular BKT transition, driven by the atomic-pair field dynamics, 
does not satisfy the  normal BKT scaling $C_m(T_{BKT}) \propto L^{-1/4}$ but 
satisfies the scaling of the atom-pair  such that $C_m(T_{BKT}) \propto L^{-1}$ as shown in 
Fig.~\ref{Scaling_atomique_molecular_condensate_vs_Temperature} (b).
This result is in good agreement with a previous study of an effective $XY$ coupled model \cite{deForges2016}.
Therefore, the fact that both atomic and molecular BKT transitions occur at the same critical temperature $ T_{BKT}/U\simeq 0.23$, see Fig.~\ref{Scaling_atomique_molecular_condensate_vs_Temperature},
indicates that the molecular field mimics the behavior of the atomic pairs exactly, or in other words $g/U=0.6$ is a strong coupling.
The order and the universality class of the thermal phase transitions of the phase diagram in 
Fig.~\ref{Thermal_Phase_Diagram_rho2} are summarized in Table~\ref{tab2}.

\begin{table}
\begin{tabular}{c c}
\hline
\hline
Thermal Phase Transition &   Type\\
\hline
 \multirow{2}*{$\rm SF_m$-NBL }     &    BKT with anomalous \\
                                                          &    $8T_{BKT}/\pi$ jump in $\rho_s$ \\
                                                          \hline
 $\rm SF_{am}$-NBL  (low temperature)&  Possible $1^{\rm st}$ order\\
\hline
 \multirow{4}*{$\rm SF_{am}$-NBL }     &   Normal atomic  BKT  \\
                                                          &       with  $\eta_a(T_{BKT})=1/4$,  \\
                                                          &       Abnormal molecular   BKT   \\
                                                          &       with  $\eta_m(T_{BKT})=1$  \\
                                                          \hline
                                                          \hline
\end{tabular}
\caption{Order and universality class of the thermal phase transitions of 
the phase diagram in Fig.~\ref{Thermal_Phase_Diagram_rho2}.  \label{tab2}}
\end{table}

\section{Conclusion}
\label{section6}

Studying numerically a coherently coupled 2D atom-molecule mixture  at zero and finite temperature, we 
unveiled the phase diagram and the universal traits of the transitions. 
At zero temperature, we have shown that an insulating phase is stabilized close to the Feshbach resonance -- the Feshbach insulator -- 
by the atom-molecule conversion term  in a region where interactions alone cannot stabilize a Mott insulator.
The Feshbach insulator involved noninteger density plateaus for both atomic and molecular species such that 
$\rho_a \sim 2 \rho_m$ close to the resonance. Such a measurement,
directly accessible using Stern-Gerlach separation  during  the  cloud  expansion \cite{Herbigetal2003}, 
will be a definitive evidence that this phase is not a standard Mott phase with integer density.
The ground state phase diagram comprises the FI phase close to the resonance, a molecular condensate for negative detuning, 
and a mixed atomic-molecular condensate for positive detuning.
The richness of the phase diagram also comes from the variety of quantum phase transitions:
the transition from molecular to mixed condensate is found to be of the 3D Ising type due to the breaking of the $\mathbb{Z}_2$ symmetry associated with the 
phase of the atomic field; the transition from molecular condensate to Feshbach insulator  belongs to 
the universality class of the 3D $XY$ model; and interestingly enough, 
the transition from mixed condensate to disordered phase (vacuum or Feshbach insulator) associated with the spontaneous symmetry breaking 
of both U(1) and $\mathbb{Z}_2$  is systematically found to be of the first order; otherwise the transitions are second order.
The thermal effects are also discussed. The conversion term couples coherently and asymmetrically the phase of the atomic and molecular fields, and therefore strongly 
affects the BKT transition. This leads to an unusual molecular superfluid to normal Bose liquid BKT transition, involving a renormalized $8T_{BKT}/\pi$  stiffness jump,
instead of the standard $2 T_{BKT}/\pi$ one for the single component case.  
The transition from mixed superfluid to normal Bose liquid also requires a careful treatment since only the 
atomic BKT transition is conventional whereas the thermal disintegration of the molecular superfluid  
satisfies the scaling of the atom-pair  such that $C_m(T_{BKT}) \propto L^{-1}$. 
Finally, we observe a discontinuity in the superfluid density at the mixed superfluid to normal Bose liquid transition, 
indicating the existence of a  possible classical first-order transition.
These rich phenomena are amenable to experimental verification using state-of-the-art setups in cold-atom physics.\\

\emph{Acknowledgements.} 
We thank T. Roscilde, F. H\'ebert,  and A. Ran\c{c}on for useful discussions
and for their critical reading of the manuscript. We also thank Professor Min-Fong Yang  for  critical comments 
and suggestions.
L.dF.dP also thanks Sasha de Forges de Parny  and Solenne Ghintran for their support.
This work is supported by Agence Nationale de la Recherche (``ArtiQ" project) and the Alexander von Humboldt-Foundation. 
All calculations have been performed on the PSMN center of the ENS-Lyon.

\end{document}